\documentclass{article}

\usepackage{arxiv}

\usepackage{eurosym}
\usepackage{graphicx}
\usepackage{color}
\usepackage{amssymb}
\usepackage[ruled,vlined]{algorithm2e}
\usepackage{subcaption}
\usepackage[breaklinks=true]{hyperref}
\usepackage{amsmath}
\usepackage[autostyle]{csquotes}
\usepackage{xargs}                      
\usepackage[colorinlistoftodos,prependcaption]{todonotes}
\usepackage[inline]{enumitem}

\usepackage{smartdiagram}
\usepackage{comment}

\title{A survey of recommender systems for energy efficiency in buildings: Principles, challenges and prospects}

\author{
  Yassine Himeur\thanks{Information Fusion, 2021} , Abdullah Alsalemi, Ayman Al-Kababji, Faycal Bensaali\\
  Department of Electrical Engineering\\
  Qatar University\\
  Doha, Qatar \\
  \texttt{yassine.himeur@qu.edu.qa;a.alsalemi@qu.edu.qa;aa1405810@qu.edu.qa;f.bensaali@qu.edu.qa} \\
   \And
 Abbes Amira \\
  Institute of Artificial Intelligence\\
  De Montfort University\\
  Leicester, United Kingdom \\
  \texttt{abbes.amira@dmu.ac.uk} \\
   \And
 Christos Sardianos, George Dimitrakopoulos, Iraklis Varlamis \\
  Department of Informatics and Telematics\\
  Harokopio University of Athens\\
  Athens, Greece \\
  \texttt{sardianos@hua.gr;gdimitra@hua.gr;varlamis@hua.gr} \\
}

\begin{document}
\maketitle

\begin{abstract}
Recommender systems have significantly developed in recent years in parallel with the witnessed advancements in both internet of things (IoT) and artificial intelligence (AI) technologies. Accordingly, as a consequence of IoT and AI, multiple forms of data are incorporated in these systems, e.g. social, implicit, local and personal information, which can help in improving recommender systems' performance and widen their applicability to traverse different disciplines. On the other side, energy efficiency in the building sector is becoming a hot research topic, in which recommender systems play a major role by promoting energy saving behavior and reducing carbon emissions. However, the deployment of the recommendation frameworks in buildings still needs more investigations to identify the current challenges and issues, where their solutions are the keys to enable the pervasiveness of research findings, and therefore, ensure a large-scale adoption of this technology. Accordingly, this paper presents, to the best of the authors' knowledge, the first timely and comprehensive reference for energy-efficiency recommendation systems through (i) surveying existing recommender systems for energy saving in buildings; (ii) discussing their evolution; (iii) providing an original taxonomy of these systems based on specified criteria, including the nature of the recommender engine, its objective, computing platforms, evaluation metrics and incentive measures; and (iv) conducting an in-depth, critical analysis to identify their limitations and unsolved issues. The derived challenges and areas of future implementation could effectively guide the energy research community to improve the energy-efficiency in buildings and reduce the cost of developed recommender systems-based solutions.
\end{abstract}


\section{Introduction} 
The use of energy has been increasing exponentially through the last few years around the world. Specifically, the building sector alone consumes more than 40\% of the global energy produced worldwide \cite{cao2016building,Himeur2020AE}. This consumption is expected to increase by 1.3\% per year on average from 2018 to 2050 in organization for economic cooperation and development (OECD) countries (e.g. USA, Europe, Australia, etc.), while this rate will be more than 2\% for non-OECD countries (e.g. Middle East, China, Russia, etc.) \cite{economidou2020review}. Therefore, experts naturally assume that the rise of population and quality of life in various regions will result in a growing need for electricity-consuming devices and individualized equipment, and hence, an increasing energy consumption rate \cite{pylsy2020buildings,Himeur2020iscas}.

In order to alleviate this issue, recent research and development projects and initiatives have been focused on developing nearly zero energy buildings (nZEB) in the last decade, which incorporate renewable and sustainable energy resources and energy management systems \cite{refat2020prospect}. However, these kind of measures could not be supported in all countries around the globe due to its high deployment cost \cite{huang2018uncertainty,lin2020towards}. Consequently, finding other cost-effective or no-cost energy saving solutions became the core of interest for the building energy community, especially those based on the use of information and communication technologies (ICT) \cite{Himeur2020IJIS-NILM-R}. One of these challenging approaches is behavioral change, which allows end-users to polish their energy consumption behaviors and trim their wasted energy without investing more time and elbow grease, but only by using recommender algorithms, artificial intelligence (AI) tools and already used smartphones \cite{azizi2019making,staddon2016intervening,belhadi65deep}. To this regard, energy providers, policy makers and end-users in the building sector have become progressively aware of the importance of behavioral change in promoting energy saving and reducing carbon emissions \cite{fraternali2017encompass,casals2017serious}. In this context, an increasing number of literature, projects and commercial products have recently arisen to explore the research interest of sustainable behavior change, explicitly to address the relation between attitudes in order to improve energy consumption behavior \cite{Sardianos2020GreenCom}. This is also due to the widespread use of AI, Internet of things (IoT) devices and other ICT tools, which have a positive impact on raising end-users' awareness, shaping their attitudes towards energy saving and boosting their achievements \cite{hwang2016efficient,Himeur2020icict}.


While most of the research efforts have been conducted towards developing and improving new technologies and materials that reduce wasted energy and promote energy saving, human-related aspects, especially those related to end-user's behavior \cite{Himeur2020IntelliSys} have received less attention. Therefore, strategies and objectives must be set in order to shape the behaviors of buildings' end-users and owners \cite{becchio2018impact}. This can be achieved through developing context-aware recommender systems \cite{ashouri2018development} that combine the knowledge of AI, behavioral analytics and human decision-making processes, to implement powerful behavioral change support systems \cite{csimcsek2016semantic,Varlamis2020CCIS}, in which recommendations could be easily embedded into daily behaviors to reach an effective energy saving level. In this regard, changing daily behavior of end-users has become a key challenge \cite{iweka2019energy,sardianos2020emergence}. This challenge requires training and awareness exercises, incentive recommendations and feedback assessments for inducing a permanent change.

Despite the success of recommender systems in different research and development applications (e.g. in healthcare, online shopping, movies, music, travel plans, etc.), there is still room for more research to improve their performance, especially in the energy saving domain. According to our knowledge, no work yet has been dedicated to the survey of challenges, difficulties and future perspectives of energy saving recommender systems. As a result, there are still open questions, in the energy research community, about the reliability and efficacy of recommendation systems. To alleviate theses issues, we provide this survey article that performs an in-depth, critical analysis of energy saving recommender systems for buildings. More specifically, this paper identifies the open issues and critical challenges that impede the development of effective energy recommender systems that incorporate human-in-the-loop, by promoting energy efficiency behaviors and reducing carbon emissions. 

To that end, a taxonomy of energy recommender systems is presented. The currently available frameworks are described and the factors that impact the efficacy of current implementation are discussed. Such factors refer to the nature of developed recommender algorithms, the computing platforms used to implement them, their objectives, the incentive measures utilized to motivate end-users and the evaluation metrics that fit the context of energy. Moving forward, a critical analysis and discussion is conducted to identify the limitations and difficulties encountered when developing energy recommender systems. Finally, we derive and decipher the issues that remain unresolved and attract an increasing research interest along with the hottest research directions for recommendation systems' performance improvement. To summarize, the main contribution axes of the paper could be outlined as follows:
\begin{itemize} 
\item Propose the first review framework of recommender systems for energy efficiency in buildings.

\item Conduct a novel taxonomy of existing energy efficiency recommender systems through analyzing the nature of the different components used to build a recommender framework, e.g. (i) objective of the recommendations, (ii) methodology and algorithm of choice utilized by the recommender engine, (iii) computing platforms, and (iv) evaluation metrics and incentive measures.

\item Perform in-depth, critical analysis of the existing frameworks to identify the current challenges and difficulties that remain unresolved.

\item Provide insights about the future orientations that could be targeted to overcome existing energy efficiency recommender systems' issues, to improve their quality and facilitate their applicability.

\end{itemize}

The remainder of this paper is organized as follows. Section \ref{sec:methodology} briefly explains the methodology that we followed in order to identify the works related to the development of energy saving recommender systems. Section \ref{sec:related_work} begins with an introduction to recommender systems, and then focuses on the analysis of human-driven energy efficiency frameworks of buildings, using recommendation systems for triggering and maintaining this behavioral change. Additionally, it summarizes the objectives of such systems, methodologies and algorithms they use, computing platforms and evaluation metrics. Moving forward, Section \ref{sec4} conducts in-depth and critical analyses of existing energy recommender systems by discussing their limitations and issues. Following, Section \ref{sec5} presents current challenges and future orientations that should attract the attention of R\&D communities in the near and far future. Lastly, Section \ref{sec6} derives the final conclusions.

\section{Methodology} \label{sec:methodology}
In order to perform our literature review, we based our methodology on the techniques presented in \cite{kitchenham2004procedures}.
Identifying the need for a review is of equal importance to the results of the review. The need for this review derives from the fact that lots of systems and solutions have been developed in the field of energy efficiency for buildings, which add to the variety of the research field. Our study reveals that there is not yet a systematic review that explains all the steps: from the conception of an energy saving solution, to the delivery to end-users. By conducting this review, the following questions will be answered:
\begin{enumerate}
    \item Why have recommender systems gained significant attention for energy efficiency in buildings?
    \item What are the main research directions that existing energy saving recommender system frameworks followed?
    \item What are the main objectives and which methodologies were used to achieve them?
\end{enumerate}

We performed our bibliometric research under the perspective of a narrative review. Studies related to the use of recommendations for improving energy efficiency and promoting energy savings in buildings have been searched. Our search took place through the Scopus database from 2000 to 2020. The following terms have been searched in titles, abstracts and keywords: \enquote{recommendation}, \enquote{recommenders}, \enquote{recommender systems}, \enquote{energy saving}, \enquote{energy efficiency}, \enquote{buildings}, \enquote{behaviour}.
 
\begin{figure*}[!htb]
\centering
\includegraphics[width=0.8\columnwidth]{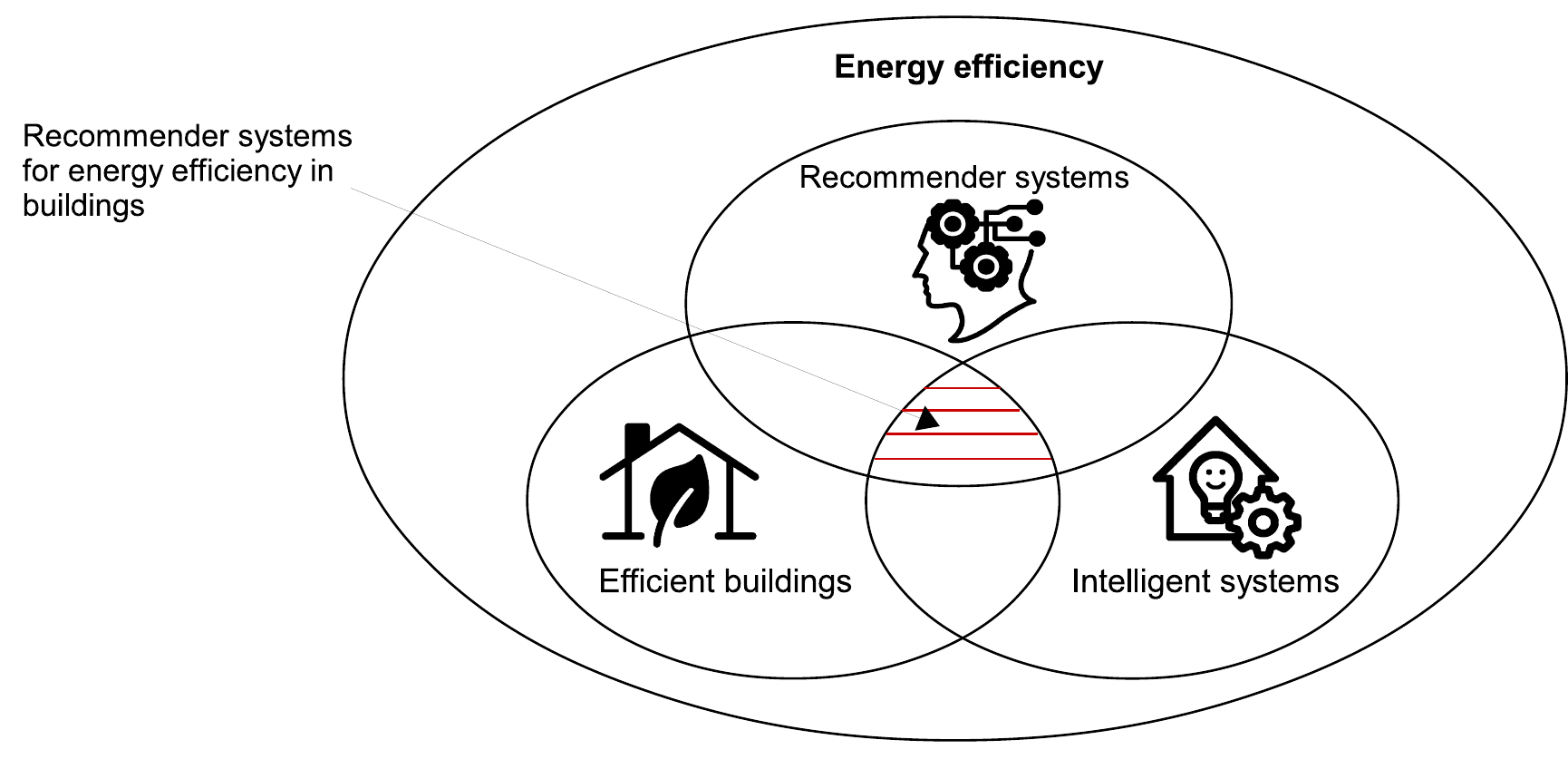}
\caption{The impact of energy efficiency in buildings in bringing human-in-the-loop.}
\label{fig:domain_survey} 
\end{figure*}

A \textbf{\textit{search in Scopus}}, in the title, abstract and keyword fields \footnote{TITLE-ABS-KEY ( ( \enquote{energy saving}  OR  \enquote{energy efficiency}  OR  energy )  AND  ( \enquote{recommendation}  OR  \enquote{recommender}  OR  \enquote{recommender systems}  OR  \enquote{recommendation systems} )  AND  behaviour  AND  buildings )} returned \textbf{\textit{283 articles}}, which are broadly organized in \textbf{\textit{three major research directions}}, as we explain in the following paragraphs and depict in Fig.~\ref{fig:domain_survey}:
\begin{enumerate}
    \item Recommendations for enhancing buildings energy efficiency
    \item Intelligent systems that promote energy saving in buildings
    \item Recommender systems that put humans in the center of the decision making process for energy efficiency, in each of the previous cases or in both
\end{enumerate}

A large group of works focuses on energy efficient buildings by design. Such \enquote{high-performance} buildings employ energy optimization techniques by including natural ventilation, thermal storage and optimal window size and placement. The number of works in this field is vast, but most of them are slightly related to recommender systems and human behavior. So we suggest readers to consult a few survey works in this domain \cite{hauge2011user, taherahmadi2020toward}.

Another major group of works focuses on the use of intelligent systems for monitoring and reducing the buildings' unnecessary energy waste. The survey of Boodi et al. \cite{boodi2018intelligent} summarizes the state-of-the-art works in building energy management system (BEMS) and distinguishes three types of models that combine environmental conditions, energy prices, comfort criteria and occupancy prediction in order to optimize the heating, ventilation and air conditioning (HVAC) systems operation: white box, black box and gray box models. More work on intelligent systems for energy efficiency in buildings can be found in a few more surveys \cite{de2014intelligent, khajenasiri2017review, shareef2018review}.

The third group of works employ recommendation systems and algorithms, usually as a complement to the previous two approaches (i.e. on energy efficiency and smart buildings). The final decision is always on the human, who plays a vital role in the efficacy of the proposed solution. The recommendations in this group are either targeted to building owners (mainly for large public or commercial buildings) or to the occupants of residential buildings. The main difference between the two, is that in the former case, recommendations refer to an energy plan or an energy saving strategy that can be used to balance between saving and comfort, whereas in the latter one, the recommendations are about energy saving actions (e.g. device turn-off, or work shifting) that can have an immediate impact on the building's energy consumption. 

As depicted in Fig.~\ref{fig:domain_survey}, the focus of our survey is on recommender systems for energy efficiency and more specifically on the intersection of the three aforementioned domains. In the following, we examine various aspects of systems that combine intelligent systems and action recommendations to improve energy efficiency and convert conventional buildings into smart ones. The section that follows begins with an overview of recommender systems and answers the last question of our survey methodology by presenting the objectives and methodologies used to achieve them.

\section{Recommender systems for energy efficiency in buildings} \label{sec:related_work}

According to the
2012 ACM Computing Classification System\footnote{https://www.acm.org/publications/class-2012}, recommender systems are categorized as information systems that focus on information retrieval tasks. Under this prism, it is now very common for more and more scenario-specific applications to adopt various kinds of recommender systems to serve their needs and goals. Although they have originally been applied online for content personalization based on users' explicit or implicit preferences \cite{resnick1997recommender, schafer1999recommender}, they soon have been extended to a wide range of different real-world applications \cite{martin2009recsys}, from place and people recommendations based on location \cite{bao2015recommendations} to action recommendations for reshaping energy profiles \cite{alsalemi2020achieving, sardianos2019want}. In addition, due to the high demand of personalization in most of real-life scenarios, various approaches of adopting recommender systems have been proposed based on the type of decision making the system has to support and the goal these recommendations have to meet. Although all these approaches still share the same logic as before (i.e. recommend items to users), they go beyond content personalization and consequently increase the requirements from recommended item, which apart from matching a specific user's interests, also have to be novel, profitable, feasible in terms of the user context, generated at the right place and moment \cite{eirinaki2018recommender}.

\begin{figure*}[!htb]
\centering
\includegraphics[width=0.8\columnwidth]{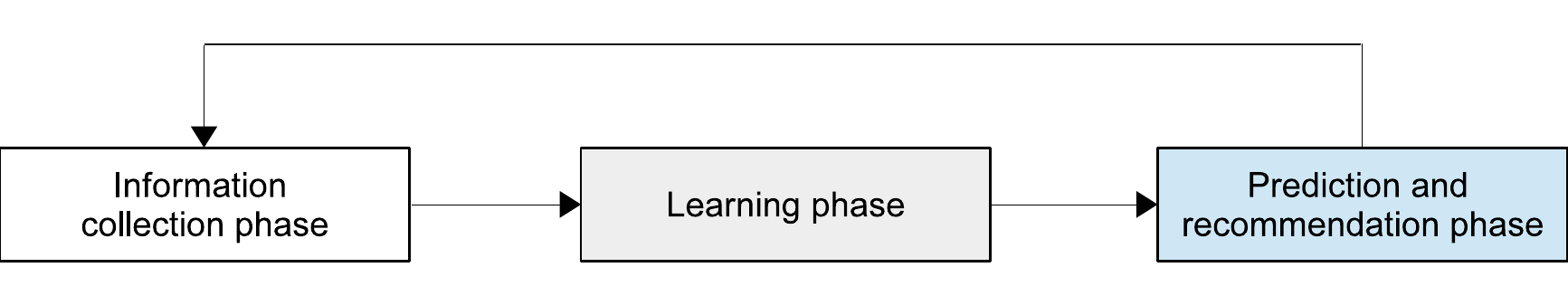}
\caption{Phases of recommendation process.}
\label{rec_phases} 
\end{figure*}

The used approach for implementing a recommender system significantly depends on the goal and applicability of the system. As depicted in Fig.~\ref{rec_phases}, the main phases that each recommendation process consists of are: 
\begin {enumerate} 
\item \textbf{Information collection phase}:
In this phase, all the necessary explicit or implicit user information is collected in order to profile the targeted users. Depending on the type of the used recommender system, the information needed for user profiling might differ in volume or attributes, but in minimum it has to cover the case scenario for which the recommender system is used. It is critical, in this phase, to recognize and prepare this information, which will be used for system training and fine-tuning in the next phase.
The efficiency and impact of the resulting recommendations rely on the utilized algorithm, but also depend on the quality of training data.
\item \textbf{Learning phase}:
In this phase, the system extracts the most representative features and trains the model that best identifies and quantifies the relationship among the users and the \enquote{items} that the recommendation engine will create recommendations for.
\item \textbf{Prediction and recommendation phase}:
In the third and last phase of the recommendation process, the system predicts the \enquote{unknown} values of the user-to-items preferences using the pre-trained model, and ranks the items that are most likely to fit users' preferences. As a result, it adds the top ranked items in the list of recommendations that are to be presented to the user. Of course, filters can be employed to rule out items that do not match the user context, and more criteria can be used to improve the ranking of items to be recommended \cite{castells2015novelty}.
\end {enumerate}

\begin{figure*}[!t]
\centering
\includegraphics[width=\columnwidth]{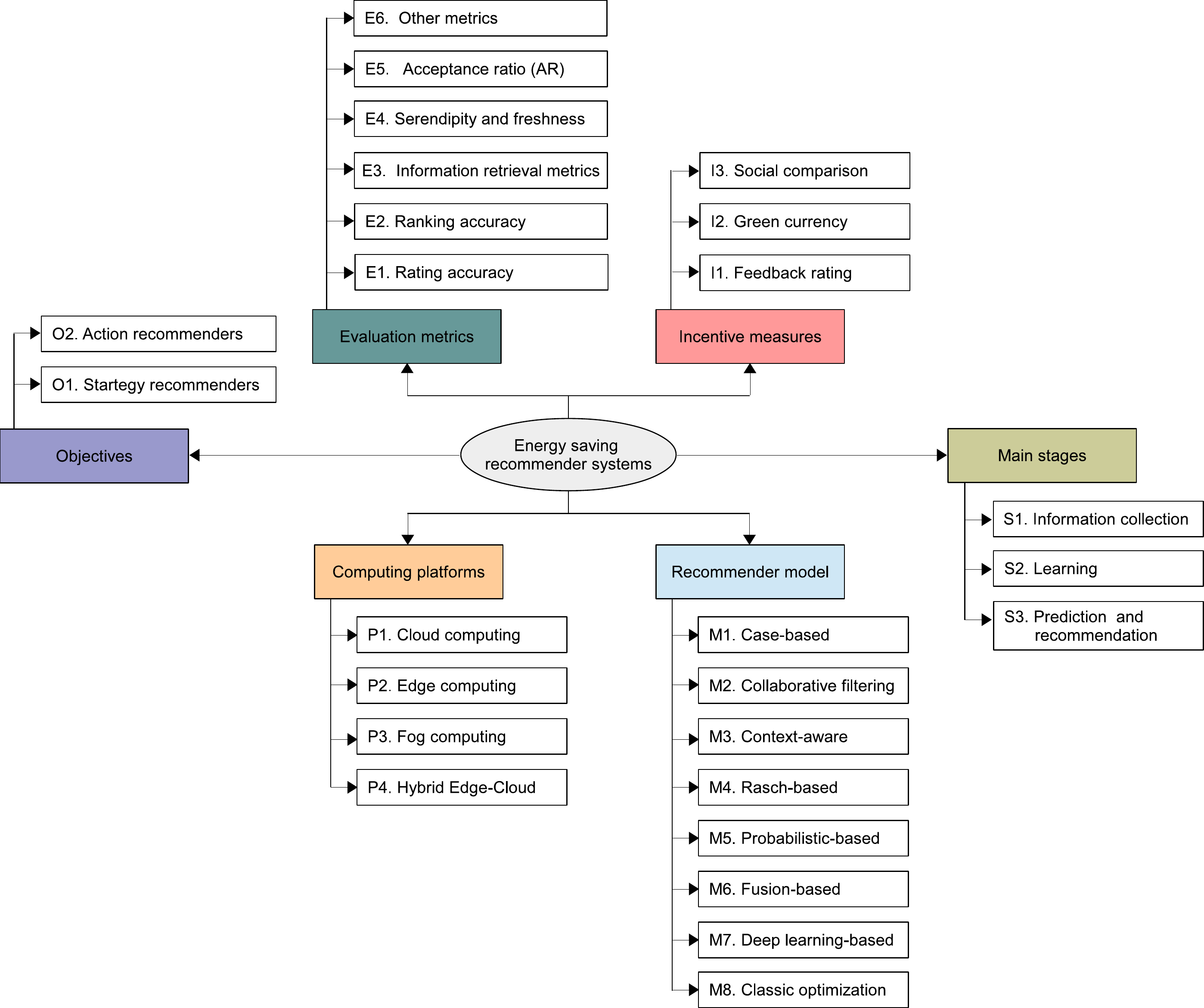}
\caption{Taxonomy of energy saving recommender systems.}
\label{taxonomy} 
\end{figure*}

In following sections, we present a taxonomy of energy saving recommender systems, in which we describe state-of-the-art research frameworks with the aim of identifying the most prominent and recent advances of this technology. Fig.~\ref{taxonomy} illustrates a taxonomy of energy saving recommender systems that is introduced based on different parameters, including the objective of the recommender system, the recommendation methodology, the computing platforms, the evaluation metrics, and the employed measures to encourage end-users to adopt energy saving behavior.

\subsection{The objectives of energy efficiency recommender systems}

The analysis performed in this article 
focuses on the third group of research works that we identified through the literature survey. As we explained earlier, the group comprises two main subgroups, each one targeting a different audience, and thus, having different objectives.

\textbf{O1. Strategy recommenders:}
The objective of such systems is to \textbf{\textit{recommend the best strategy}} for each case, either it is a building, an energy consumption forecast model, or an operational energy setup for HVAC or lights. For example, \cite{Pinto2018} proposes a building energy efficiency recommender system that relies on case-based reasoning (CBR). 
In a similar direction, \cite{kaur2019energy} discusses a recommender system for selecting the operational light intensity that satisfies user's comfort that is suitable for both user activity and energy saving.
Finally, \cite{cui2016short} presents a generalized structure for forecasting building energy trends based on building specifications. The solution is mostly targeted to energy prosumers and allows them to predict the total energy consumption of a building by choosing the right building profile.

\textbf{O2. Action recommenders:}
These systems are tailored to the everyday needs of building occupants and their main objective is to \textbf{\textit{recommend actions that minimise the energy footprint of occupants}}. The actions can either assume static users and inelastic needs, or build on their flexibility to move around the building or postpone their needs at a later time (e.g. using the laundry machine after hours). For example, 
\cite{cuffaro2017resource} develops a Resource-oriented rule-based engine that generates advice for energy savings in the form of real-time alerts or logged incidents for monitoring purposes. On the other side, \cite{Wei2018} focuses on commercial buildings and distinguishes two types of recommendations, which both aim to maximize the space usage: recommendations for occupants to move from one space to another, and recommendations for occupants to shift their schedule related to the building. In a more recent work \cite{wei2020deep}, authors present a recommender system for reducing energy consumption in commercial buildings. They employ human-in-the-loop methodologies and utilize deep reinforcement learning in order to learn actions with energy saving capability and actively deliver recommendations to building end-users.
Once again, the recommendations have to balance between user comfort and energy efficiency, but the final decision lies in the end-users' hands. For example, ReViCEE \cite{KAR2019135} analyzes historic energy usage fingerprints and provides individual and collaborative recommendations that balance between comfort and efficient usage of energy.


\subsection{Methodologies and algorithms for energy efficiency recommendations} \label{sec3.2}
In this section, we overview recommendation systems methodologies widely used in the building energy saving field.

\vskip3mm

\noindent \textbf{M1. Case-based} recommender systems are in essence rule-based systems, which recommend actions for one or more end-users through handling every end-user separately. Explicitly, specific energy usage habits and preferences are examined with reference to an ensemble of rules/heuristics and predetermined decisions that initiate, if achieved, the associated energy efficiency acts.

In \cite{Schweizer7424470}, a rule-based recommender system is implemented to effectively learn the energy consumption patterns and interests of end-users, and hence, allows them to independently promote energy efficiency. In this regard, a frequent-sequential data mining model is deployed to extract the characteristics of consumers. Similarly, in \cite{OSADCHIY2019535}, a pairwise association rule-based algorithm is proposed for drawing the collective preferences of groups of end-users. The resulting recommender system is able to provide personalised suggestions to users without requiring a complex rating approach. While in \cite{Pinto2018}, the authors propose a case-based reasoning recommendation system, which is built based on system knowledge (i.e. cases) representing the historical energy consumption actions in order to shape the end-user behavior towards an energy efficiency comportment. Explicitly, the system is able to suggest energy efficiency actions to end-users at every different moment of the day. This is done by analyzing their energy usage footprints and comparing them with the ones already stored in the knowledge base. In this line, in order to identify the similar behaviors at every time-stamp, a k-Nearest Neighbor (KNN) approach is deployed, while a support vector machine (SVM)-based method is utilized to optimize the weighting parameters of every example. Moving forward, an expert system is then employed, which includes a set of ad-hoc rules for ensuring the application of the developed scheme to the case under consideration. Recently, in \cite{dahihande2020reducing}, pattern mining techniques are used for creating appliance usage profiles that consider user time context. Then, they filter out the most prominent time-appliance patterns that indicate the ideal behavior and when they detect an appliance usage that deviates from the typical behavior, they send a recommendation notification to the end-user to interact accordingly with the device.

\vskip3mm
\noindent \textbf{M2. Collaborative filtering} methods assume that the ensemble of end-users select from a closed set of actions (or items). These explicit or implicit choices are used to filter (predict) the preference of any end-user for any action in the set of available actions (or items). 
Therefore, the recommended actions to each end-user represent the most preferred to him/her (or those having the highest estimated rating) \cite{morawski2017fuzzy} or those generated from the same group he/she pertains to \cite{castro2018group}. In this line, energy efficiency recommender systems, using collaborative filtering, utilize various intelligent agents, which interact proficiently and dynamically identify the end-users' preferences. This helps in promoting energy saving actions to consumers by providing them with personalized recommendations that appropriately fit their interests.

For example, in \cite{Zhang8412100}, appliances' consumption data of a specific household are analyzed before predicting the rating levels of different energy usage plans and identifying the related user preferences for every plan. Moving forward, a filtering (prediction) model is used to allow users to choose suitable consumption plans with adequate tariffs. In the same way, the authors in \cite{ZHENG2020117775} opt to generate energy efficiency recommendations based on a dual-step procedure, i.e. extracting features and triggering tailored recommendations. Explicitly, at the first stage, a matrix representation is adopted to adapt user preferences with reference to appliances usage. Following, a collaborative filtering approach is employed to detect similar users and generate personalized recommended actions using KNN clustering. While in \cite{KAR2019135}, the authors design the ReViCEE recommender system, which delivers tailored recommendations to end-users at a university campus building in Singapore, helping them curtail their electricity usage. Accordingly, ReViCEE learns end-users' interests by analyzing historic energy usage fingerprints. In this regard, individual and collaborative preferences related to the use of lights are captured from current consumption patterns before triggering a set of recommendations to grant end-users the best compromise between visual comfort and energy saving.

\vskip3mm
\noindent \textbf{M3. Context-aware} recommender systems aim to produce more pertinent recommendations which are adjusted to the particular contextual circumstances of the end-user \cite{Adomavicius2015}. This can be based on the detection of historic energy consumption patterns and their underlying context, which can then be used to develop rule-based recommendations aiming to achieve end-user's satisfaction \cite{sardianos2020model,RAZA201984}. They usually require a longer interaction with end-users with the system in order to collect a larger amount of data that helps in a better adaptation of the generated recommendations to the specific contextual situation of the consumer \cite{alsalemi2020micro}. 

The authors in \cite{Shigeyoshi2013} introduce a context-aware recommendation system based on (i) the analysis of power consumption footprints in various contexts; (ii) the maintenance of a base of historic formulated recommendations to avoid replicated recommendations; and (iii) a social survey for evaluating the recommendations efficiency, in which 47 users adopted the recommender system suggestions and provided feedback on their efficiency. The empirical assessment illustrates that when the recommendations are chosen randomly and in large numbers, they can overflow users and may have a detrimental impact on the end-users.

\vskip3mm
\noindent \textbf{M4. Rasch-based} models represent a psychometric paradigm that is generally utilized to analyze the user's responses to a specific set of recommendations \cite{radha2016lifestyle}. It aims at identifying the best compromise between the user's behavior and the ability to implement the generated recommendations. In this line, a Rasch-based recommender system relies on performing a Rasch analysis, which explains the probability of the end-user to perform a particular recommendation as a function of the end-user's ability and the recommendation’s difficulty \cite{Starke2015}. 

For example, the authors in \cite{Starke2017} investigate to what extent Rasch-based recommendations could help in reducing the end-user' efforts, improving the system assistance while increasing the choice satisfaction and leading to the promotion of energy efficiency behaviors in buildings. To this end, a Rasch-based recommendation system is developed, where up to 79 energy-efficiency recommendations are generated to assist end-users in making correct energy usage actions, enhancing system support, collecting their feedback and rating their satisfaction. 

\vskip3mm
\noindent \textbf{M5. Probabilistic relational models (PRM)} can be used to capture the knowledge hidden in the energy consumption data in a probabilistic manner, which represents the probability of an action to match certain usage patterns or preferences
\cite{chulyadyo2014personalized}. PRMs replace the user-item preference matrix with a relational database with \enquote{Users} and \enquote{Items} being the main entities and real transactions being the captured relationship between the two. When a transaction between a user and an item is recorded, the probabilities for users and items with similar attribute values increase. 
In this essence, probabilistic relational paradigms are developed for predicting end-user's preferences and habits, where tailored recommendations are derived to motivate end-users to reduce their wasted energy \cite{kumar2001recommendation}.

In \cite{Li7093924}, the authors record and analyze historical energy usage footprints in a work space via the use of a continuous Markov chain model. The model focuses on investigating time-series energy data using a multi-objective programming scheme. Next, a tailored energy usage advice is drawn and each end-user is notified to establish energy saving measures and support the adoption of renewable energy solutions. Similarly, in \cite{wei2020deep}, a recommender system aiming at optimizing power usage in commercial buildings is introduced using human-in-the-loop model. Moving forward, the energy saving approach for a building is implemented and modeled based on a Markov decision paradigm, and a deep reinforcement learning scheme is implemented to learn energy saving recommendations, and engage consumers in effective sustainable behaviors. The implemented system helps in learning end-users' energy consumption actions/habits with a good accuracy, and hence, results in providing end-users with useful and personalized recommendations. Finally, further experiments are conducted to employ a feedback rating procedure to evaluate user satisfaction and identify the best recommendations.

\vskip3mm
\noindent \textbf{M6. Fusion-based} models rely on the analysis of different kinds of data, such as energy consumption footprints, ambient conditions (i.e. temperature, humidity, light, and luminosity), outdoor weather information and user preferences/habits, for producing better and well-timed recommendations \cite{OKU2011}. The so-called fusion-based recommendation systems adopt data fusion approaches, which either collect and analyze different kinds of data representations from distinct sources before making decisions \cite{xin2011effective}
or include various sub-recommenders and aggregate their recommendation outputs, thus, building a recommendation ensemble \cite{zhang2010fusion,himeur2020data}.

For example, in \cite{wroblewska2020multimodal}, authors introduce a multi-modal embedding fusion-based recommendation system that combines information from multiple sources and modalities. In \cite{ji2020brs}, a hybrid recommender system is proposed, which is based on the assumption that the end-user choice is generally impacted by its direct (and even indirect) friends' preferences. The system fuses different kinds of data (e.g. social data, score, and review patterns) and trains a preference prediction model, using a joint-representation learning process, to extract the best recommendations. 
In \cite{shambour2012trust}, authors incorporate additional information from the consumer's social trust network as well as actionable semantic-domain knowledge, in order to improve the recommendations accuracy and increase their coverage.
In the same manner, in \cite{wang2019collaborative}, by exploiting different social data sources (produced by the Internet, e.g. consumer profiles, social relationships, behaviors, preferences, etc.), a recommendation system using social data fusion is proposed. Explicitly, it aims to utilize social data fusion for identifying similar consumers, and hence, updating each consumer rating of recommended actions using similar consumers.

In \cite{pradhan2020multi}, a multi-level fusion-based recommender system is developed to produce collaborator recommendations. It fuses deep learning and biased random walk models to provide tailored recommendations for possible end-users having the same preferences. Following, in \cite{aiello2018decision}, a recommender system is introduced to support sustainable greenhouse management in buildings using multi-sensor data aggregation based system. Specifically, contextual information, mathematical formulations and experts' knowledge are used and fused to help in generating more effective recommendations.

\vskip3mm
\noindent \textbf{M7. Deep learning-based} recommender systems, have gained significant attention recently in various research topics, including visual recognition, healthcare, fraud detection, natural language processing, etc. \cite{HimeurCOGN2020}. Their use is extended due to their remarkable performance in many learning tasks, and additionally because of their interesting ability to learn characteristic representations from the ground up. The impact of deep learning is widespread as well to other research issues, in which it demonstrates its efficiency to retrieve information and trigger recommendations. Evidently, the field of deep learning in recommender system is flourishing \cite{Zhang10.1145/3285029,app10072441}.

In \cite{R2020113054}, with the aim of addressing the gap in collaborative filtering-based methods, a deep learning model is adopted. In effect, collaborative filtering systems are seriously suffering from the cold start issue, especially with the absence of historic data about the users and their energy consumption preferences. Moreover, the latent parameters learnt by these systems are naturally linear. To that end, deep learning is employed, where embeddings are deployed to represent users and their preferences, and thus, to allow the learning of non-linear latent parameters. This approach better alleviates the cold start issue, since information about users and their preferences is embedded in the deep learning model. 
In \cite{app10144926}, a collaborative filtering approach, called DeepMF, which combines deep neural networks with matrix factorization is introduced for improving both the predictions of end-users' preferences and the provided recommendations. In this context, DeepMF applies an iterative refinement of a matrix factorization paradigm based on multilayer architecture, in which the acquired knowledge from a layer is used sequentially as input for the following layer, and so on. 

\vskip3mm

\noindent \textbf{M8. Classical optimization} is proposed in addition to learning-based recommender systems to achieve an optimal energy usage in households and other kinds of buildings. Indeed, an essential part of the literature review is based on classical optimization techniques, which can (i) provide relevant recommendations to both end-users and energy providers; and (ii) reduce wasted energy automatically through controlling energy demand and electrical devices \cite{shah2019review,tushar2019optimizing}. For example, in \cite{ceballos2019simulation}, the authors propose an energy optimization approach, which aims to predict the amount of energy used by the heating and cooling systems in a set of commercial or institutional buildings. Following, the potential impacts of various energy saving measures based on parameter optimization are investigated before recommending tailored actions to optimize energy consumption. Similarly, in \cite{rocha2015improving}, Rocha et al. discuss the potential of energy saving in buildings using efficient energy and policy measures. Accordingly, an optimization model that mimics a smart building energy management system is developed through the aggregation of the decisions on heating and cooling systems operations with decisions on energy demand. Moving forward, an optimization and ontology-driven multi-agent recommender system is introduced in  \cite{anvari2017multi} to reduce wasted energy. It can monitor and optimize energy usage within an integrated home/building and/or microgrid systems using different renewable energy resources and controllable loads.
In this regard, several agents are developed and integrated together with the aim of improving their cooperation and optimizing the operation strategy of the whole energy system.

In \cite{lu2020economic}, in order to optimize energy saving and guarantee a perfect thermal comfort of the end-users, the uncertainties due to outdoor weather conditions, building parameters and human behaviors are thoroughly modeled. Following, an adaptive economic dispatch approach is introduced, which is based on conducting a thermal comfort management process using a two-step algorithm. Similarly, in \cite{di2017two}, aiming at reducing energy demand uncertainties and energy bills in households and small public buildings, a two-step energy monitoring framework is proposed. Specifically, uncertainties due to energy demand variations and prediction errors of renewable generation are detected before generating appropriate recommendations to reduce wasted energy. Moreover, in \cite{salakij2016model,yu2017model}, energy usage optimization is conducted by forecasting the heat and moisture transfer, which directly affect the indoor climate and the overall thermal performance of buildings.

In \cite{paul2019real}, reducing wasted energy in various buildings is ensured by optimizing the consumed energy taking into consideration the abrupt changes produced by the rooftop solar generation and the real-time price of energy. Accordingly, a novel parameter called the load criticality rate has been deployed, which represents the threshold value applied by each building occupants to their power consumption. Furthermore, the energy reduction task is considered as a stochastic, multi-objective optimization issue. While in \cite{lu2019thermal}, an analytical model that describes thermal dynamic characteristics of district heating networks in buildings is developed to optimize energy consumption while keeping an acceptable comfort level of the end-users.

\subsection{Computing platforms}
Aiming at bridging the gap between development and implementation of energy efficiency recommender systems, this section presents the main computing solutions that can be utilized, as a standalone solution or in an integration of various ones, to implement these systems.

\vskip3mm
\noindent \textbf{P1. Cloud computing} relies on the use of cloud infrastructure principles to provide reliable and scalable methods to solve resource-intensive computational issues \cite{abbasi2019software}. The employment of cloud-based services and solutions in home automation and building energy monitoring has been widely and globally popular. The cloud computing model facilitates a broad variety of processing applications, exploits the potentials of IoT and leverages IoT processing restrictions by moving the most demanding parts such as deep learning algorithms \cite{AlsalemiRTDPCC2020} to the cloud. The main challenge for cloud computing solutions remains to be the privacy of IoT data \cite{zhou2017security} that are transferred and processed on the cloud platform.

\vskip3mm
\noindent \textbf{P2. Edge computing} receives increasing attention although transmitting data to the cloud for processing has become a central topic in recent decades, pushing cloud computing as a prevailing model in computing \cite{ren2019survey}. In effect, the rapid growth in the amount of devices and data traffic in the age of IoT puts major hurdles on the capacity-limited Internet and on unregulated service delays. Through utilizing cloud storage alone, it becomes impossible to fulfill the delay-critical and context-aware service specifications of IoT apps. Met with these problems, computing paradigms are moving from clustered cloud computing to dispersed edge computing. Edge computing allows some of the data processing to be done on the device (i.e. the edge) to lift some of the burden off the cloud server \cite{Alsalaemi2020sca,HIMEUR2020115872,Himeur2020icpr} and guarantee privacy, grace to the decentralised processing.

\vskip3mm
\noindent \textbf{P3. Fog computing} has been recently used for developing distributed, low-energy recommender systems based on IoT architectures. Their main applications are in the domains of healthcare \cite{devarajan2019fog}, banking \cite{hernandez2020fog}, or information brokerage \cite{wang2019fog}. Despite the many advantages of fog computing, which comprise low latency, privacy, uninterrupted service and location awareness, there is still no application that combines fog computing with energy efficient recommendations.

\vskip3mm
\noindent \textbf{P4. Hybrid edge-cloud approaches} stand between cloud and edge computing, and their primary aim is to increase the flexibility of IoT systems by moving data processing from the edge to the cloud and vice-versa, depending on the system constraints \cite{linthicum_edge_nodate}. For example, when a small dataset is at use, hybrid systems decide to offload it to the edge device for rapid processing and for reducing the communication overhead, while in more demanding situations, data are pushed to the cloud in order to get more efficient results. Also, employing an edge-cloud architecture can help distribute computations between the edge and cloud for performance optimization \cite{huang2019deepar}. This flexibility of hybrid approaches creates new, out-of-the-box possibilities for more powerful yet resource-efficient IoT solutions.

\subsection{Evaluation metrics}

Modelling a recommender system that fits the need of the business/initiative, encompasses an evaluation phase that tests the recommender's capabilities to the limits. A recommender system suggests items (or actions) to the users, based on their own expected preferences. Several metrics have been devised for this purpose that allow system performance evaluation in predicting and providing sensible recommendations that fit a given scenario. Among the long list of metrics that can be used for the evaluation of provided recommendations \cite{gunawardana2015evaluating, wu2012evaluating}, we highlight those metrics that we see relevant to an energy efficiency recommendation system.

\vskip3mm
\noindent \textbf{E1. Rating accuracy} measures the deviation between the predicted and the actual ratings assigned by a user to each recommended item. The simplest and most popular error are \textit{Mean Absolute Error (MAE)} and \textit{Root Mean Square Error (RMSE)} \cite{chai2014root}, which are defined as follows:
\begin{equation} \label{eq:accuracy}
    MAE= \dfrac{1}{|\hat{R}|}\sum_{\hat{r}_{ui}\in \hat{R}} |r_{ui}-\hat{r}_{ui}|
    \qquad \qquad 
    RMSE= \sqrt{\dfrac{1}{|\hat{R}|}\sum_{\hat{r}_{ui}\in \hat{R}} (r_{ui}-\hat{r}_{ui})^2}
\end{equation}
where $r_{ui}$ is the actual rating of user $u$ for item $i$ and $\hat{r}_{ui}$ is the predicted rating.

\vskip3mm
\noindent \textbf{E2. Ranking accuracy} assumes that items are ranked in a decreasing-rating order for each user and only the top items are presented. So, they evaluate the similarity in the order of rated items, providing a more robust evaluation method than MAE or RMSE.
Pearson ($c$) and Spearman ($\rho$) correlation coefficients measure the linear relationship between two (parametric/non-parametric) variables, as defined by the following equations:
\begin{equation} \label{eq:correlation}
    c(x,y)= \dfrac{1}{N-1}\dfrac{\sum^N_{i=1}(x_i-\bar{x})(y_i-\bar{y})}{s_x \,s_y} \qquad \qquad \rho(u,v) = \dfrac{1}{N-1}\dfrac{\sum^N_{i=1}(u_i-\bar{u})(v_i-\bar{v})}{s_u \,s_v}
\end{equation}
where $x_i$ and $y_i$ are the $i^{th}$ elements in the variables of interest, and $\bar{x}$ and $\bar{y}$ are the sample means. Similarly, $s_x$ and $s_y$ represent the sample standard deviation for a sample of size $N$. $u$ and $v$ are the ranked variables counterpart of $x$ and $y$. The values vary between -1 and 1 where the former exhibits a strong negative relation between two variables, and the latter being a positive one. A value of 0 indicates the absence of any linear relationship.

Such metrics can evaluate the recommender system as a whole, regardless of the used rating scale \cite{herlocker2004evaluating}. They enable developers to assess the recommender's ability to provide ratings for energy suggestions that are consistent with existing users' ratings, thus, evaluating the quality of the given suggestions. According to an experimental work done by \cite{herlocker2002empirical}, both Pearson and Spearman produced quite similar results, thus, being redundant if both used. Therefore, one can only use either in the context of recommender systems.



\vskip3mm
\noindent \textbf{E3. Information retrieval metrics}, such as precision, recall, and F1 score, are among the most relevant metrics that can be used in recommendation systems context with slight modifications. 
Since the item ratings are usually in a 1-5 scale, a threshold value $T$ is used to convert an absolute rating to a binary prediction that classifies whether the item is relevant or not \cite{herlocker2004evaluating}. 

The metrics are depicted in equations (\ref{eq:recall}-\ref{eq:f1_score}). In the context of recommender systems, true positive (TP) indicates that the recommended energy saving action is relevant within the user context, or has been accepted by the user, while false positive (FP) means that the recommended action is irrelevant. On the other hand, false negative (FN) indicates that the recommender system failed to recommend an energy-efficient action, that was actually performed by the user, and lastly, true negative (TN) refers to a system that does not provide energy-efficient suggestions when the context does not demand one. 
\begin{equation} \label{eq:recall} 
    Recall = \dfrac{TP}{TP+FN}
\end{equation}
\begin{equation} \label{eq:precision} 
    Precision = \dfrac{TP}{TP+FP}
\end{equation}
\begin{equation} \label{eq:f1_score}
    F1 \,\, Score = 2\times\dfrac{Precision~{\times}~Recall}{Precision + Recall} = \dfrac{2TP}{2TP+FP+FN}
\end{equation}

To elaborate further, if an excessive energy consumption is detected, multiple recommendations can be relevant to reduce the energy consumption. For instance, if a television or a room's lighting were left on, while no one is in the room, the recommender system advising to turn-off either the television or the room's lighting would be relevant (i.e. TP) to reduce household's energy consumption. On the other hand, FP would be suggesting to turn-off either the television or the lighting in a room where someone is currently within.

\vskip3mm
\noindent \textbf{E4. Serendipity and freshness} measure the novelty and variability of items recommended to the users in an effort to avoid recommending the same items again and again.
Serendipity measures the amount of surprise an accepted action generates for a user \cite{gunawardana2015evaluating}. As if the recommender system is informing the user about a new piece of information accompanying an action, which is relevant but he/she has not heard about before. From this perspective, a distance metric adapted from \cite{gunawardana2015evaluating}, defined by equation \eqref{eq:serendipity-distance}, is utilized to check the serendipitous virtue of the recommender system. It calculates the distance between a suggested action $a$ and a set of previous suggestions $A$ that the user considered. $n_{A,\,c}$ is the number of suggested actions within the same group $c \in C$ in $A$, $n_A$ is the maximum number of suggested actions from a single class $c$ in $A$, and $c(a)$ is the class of action $a$. 
\begin{equation} \label{eq:serendipity-distance}
    d(a,A) = \dfrac{1+n_A-n_{A,\,c(a)}}{1+n_A}
\end{equation}

Equation \eqref{eq:serendipity-distance} estimates the distance a recommender system accumulates for a list of suggested actions. As an example, imagine a user $i$ was subjected to a list of actions $A$ throughout a certain period. By dividing this set to two subsets of suggested observed actions $A^i_{o}$ and hidden ones $A^i_{h}$, the $A^i_{o}$ can be utilized by the recommender system to generate energy-efficient action suggestions. Now, if the recommender is asked to generate 10 actions, we would like the recommender to suggest valid and relevant suggestions to the user $i$, which are NOT in the $A^i_{o}$ set, thus, increasing the system's serendipitous virtue. This is measured by calculating the distance score $d(a,A^i_{o})$ for each $a \in A^i_{h}$, where the score will be reduced if actions from the same class as $a$, depicted in $c(a)$, are numerous, i.e. $n_{A,\,c(a)}$ is large. It is worth noting that, in a sense, serendipity battles the accuracy of the recommender system, thus, it is important to periodically check the relevance of suggested actions, as users can refrain from using the recommender system if it kept suggesting irrelevant ones \cite{gunawardana2015evaluating}.

Freshness, on the other hand, indicates the recommender capability in suggesting new recommendations each time the user interacts with \cite{mcnee2006making}. However, since the given actions to reduce energy consumption are limited by nature in small households, it is possible to adjust the definition such that the same action is not recommended in a matter of few hours/days. Thus, equation \eqref{eq:serendipity-distance} can be revisited and utilized every $k$ hours/days to ensure that an action $a$ was minimally suggested in the previous $k$ hours/days.

\vskip3mm
\noindent \textbf{E5. Acceptance ratio (AR)} 
aims to quantify the agreement that a certain user exhibits to energy efficiency suggestions provided by the recommender system. In numbers, the ratio allows the recommender system to understand the underlying probability of accepting the suggestions it provides, either on a holistic-level, or for a certain energy efficiency suggestion, e.g. turning off the air conditioner, as seen in equation \eqref{eq:acceptance_ratio}.
\begin{equation} \label{eq:acceptance_ratio}
    AR = \dfrac{1}{C}\sum_{c=1}^CAR_c\qquad \qquad AR_c = \dfrac{1}{M}\sum_{i=1}^M\dfrac{a_{i,c}}{r_{i,c}} 
\end{equation}
where $r_{i,c}$ is a recommendation and $a_{i,c}$ is either 0 or 1 depending on whether the recommendation $r_{i,c}$ is accepted. $M$ is the number of recommendations per energy efficiency suggestion belonging to the same group $C$.

The importance of this metric prevails when the recommender decides to send a suggestion, where it helps in answering the question: Which energy saving suggestion should the recommender system send, an extreme energy saving suggestion with low acceptance rate, or a moderate one with high acceptance rate?

\vskip3mm
\noindent \textbf{E6. Other metrics} aim to evaluate: i) the \textit{Coverage} of recommendations in terms of the item or user space, which means that the recommender systems must suggest all possible actions and recommend actions to all users, ii) the \textit{Confidence} of the system in its recommendations, iii) the \textit{Trust} of users to them, iv) the \textit{Utility}, v) the \textit{Risk}, vi) the \textit{Robustness} and many more.
\cite{alsalemi2020achieving}


\subsection{Incentive measures}
Aiming at increasing the acceptance ratio of energy saving recommendations, recommender systems should provide the user with incentives, to encourage and motivate them in promoting more sustainable energy behaviors. In this context, several incentives have been deployed in energy efficiency recommendation systems, such as feedback rating, green currency and social comparison. 

\vskip3mm

\noindent \textbf{I1. Feedback rating} is of significant importance, because it can show the end-users' satisfaction regarding the delivered recommendations \cite{karjalainen2011consumer,karlin2015effects}. Therefore, gathering feedback ratings helps to efficiently adapt the recommendations to the users' preferences and interests, and thereby results in a substantial contribution to deliver sustainable reductions in energy consumption.

\vskip3mm

\noindent \textbf{I2. Green currency} is tightly coupled with the rapid development and prevalent use of blockchain and cryptocurrencies. The energy research community has also inspired the creation of green and/or $CO_{2}$ coins, which can be used as motivation towards energy efficiency. For instance, in \cite{garbi2019beneffice}, end-user engagement is promoted via the adoption of monetary rewards, named $CO_{2}$ credits, in recognition of energy savings and effective achievements. 

\vskip3mm
\noindent \textbf{I3. Social comparison} has recently been adopted in various recommender systems with different applications (e.g. tourism and travels, movies, e-commerce, e-learning, healthcare, etc.). In the energy efficiency scenario, social comparison aims to motivate electricity saving in households \cite{petkov2011engaging}. This research area relies on the use of fundamental theories of social psychology which can teach us on how to encourage end-users to preserve energy by recommending individuals with better profiles as a reference \cite{jain2013can,du2016modelling}. More specifically, normalized comparison modules (that perform consumer rankings) are incorporated in the eco-feedback tools, in order to help end-users to compare their energy usage patterns with those of their peers and neighbors. The effectiveness of this incentive mechanism comes from the fact that end-users are highly influenced by engagements and rankings of their peers on social networks \cite{wemyss2019does,morley2018digitalisation}.

All in all, Table~\ref{SummaryRS} summarizes the characteristics of the recent and relevant energy efficiency recommender systems. It highlights their advantages, and outlines the main information about their taxonomy, which could be extracted from the overview conducted above.

\begin{table} [t!]
\caption{Summary of the relevant energy efficiency recommender systems proposed in the literature.}
\label{SummaryRS}
\begin{center}

\begin{tabular}{llllll}
\hline
{\small Framework} & {\small Type of RS} & {\small Advantages} & {\small %
Computing} & {\small Objective} & {\small Application} \\ 
&  &  & {\small Platform} &  & {\small Environment} \\ \hline
{\small Pinto et al. \cite{Pinto2018}} & {\small M1} & {\small Recommend
operational light intensity} & {\small P1} & {\small O1} & {\small Households%
} \\ 
&  & {\small satisfying users' comfort} &  &  &  \\ 
{\small Kaur et al. \cite{kaur2019energy}} & {\small M1} & {\small Predict
energy consumption ratings and} & {\small P2} & {\small O2} & {\small %
Households} \\ 
&  & {\small offer personalized recommendations} &  &  &  \\ 
{\small Schweizer et al. \cite{Schweizer7424470}} & {\small M1} & {\small %
Learn the energy usage habits and} & {\small P1} & {\small O1} & {\small %
Households} \\ 
&  & {\small interests of consumers} &  &  &  \\ 
{\small Zhang et al. \cite{Zhang8412100}} & {\small M2} & {\small Generate
tailored recommendations} & {\small P1} & {\small O2} & {\small Households}
\\ 
&  & {\small following predicted ratings of energy usage} &  &  &  \\ 
{\small Zheng et al. \cite{ZHENG2020117775}} & {\small M2} & {\small Provide
appliance-level consumption} & {\small P3} & {\small O2} & {\small Households%
} \\ 
&  & {\small recommendations} &  &  &  \\ 
{\small ReViCEE \cite{KAR2019135}} & {\small M2} & {\small Predict
collaborative recommendations of} & {\small P4} & {\small O1} & {\small %
Households} \\ 
&  & {\small light preferences of end-users} &  &  &  \\ 
{\small Garcia et al. \cite{Garcia2017}} & {\small M2} & {\small Produce
tailored advice on end-users'} & {\small P1} & {\small O2} & {\small %
Households} \\ 
&  & {\small activities similar scenarios} &  &  &  \\ 
{\small Shigeyoshi et al. \cite{Shigeyoshi2013}} & {\small M3} & {\small %
Produce contextual based advice with} & {\small P3} & {\small O2} & {\small %
Households} \\ 
&  & {\small social experiment ratings} &  &  &  \\ 
{\small Luo et al. \cite{Luo2017}} & {\small M3} & {\small Tailored
recommendations with textual} & {\small P1} & {\small O2} & {\small %
Households} \\ 
&  & {\small appliance advertisements} &  &  &  \\ 
{\small Wei et al. \cite{Wei2018}} & {\small M3} & {\small Provide move and
shift-schedule} & {\small P1} & {\small O1} & {\small Commercial} \\ 
&  & {\small recommendations} &  &  & {\small buildings} \\ 
{\small REHAB-C \cite{SARDIANOS2020394}} & {\small M3} & {\small Tailored
recommendations with feedback} & {\small P2, P3} & {\small O2} & {\small %
Academic} \\ 
&  & {\small on end-users' preferences} &  &  & {\small buildings} \\ 
{\small Starke et al. \cite{Starke2015}} & {\small M4} & {\small Provide
Rasch profile based recommen-} & {\small P3} & {\small O1} & {\small %
Households} \\ 
&  & {\small dations of end-users' behavior} &  &  &  \\ 
{\small Starke et al. \cite{Starke2017}} & {\small M4} & {\small Generate
Rasch profile recommendations} & {\small P3} & {\small O1} & {\small %
Households} \\ 
&  & {\small based on a social experiment} &  &  &  \\ 
{\small Li et al. \cite{Li7093924}} & {\small M5} & {\small Provide tailored
recommendations to support} & {\small P1} & {\small O1} & {\small Work spaces%
} \\ 
&  & {\small the use of renewable energy solutions} &  &  &  \\ 
{\small Wei at al. \cite{wei2020deep}} & {\small M5} & {\small Optimize
power consumption using} & {\small P1} & {\small O2} & {\small Commercial}
\\ 
&  & {\small human-in-the loop model} &  &  & {\small buildings} \\ 
{\small Bravo et al. \cite{Bravo2019}} & {\small M6} & {\small Create energy
saving recommendations} & {\small P2} & {\small O1} & {\small Households} \\ 
&  & {\small based on electricity price} &  &  &  \\ 
{\small Aiello\ et al. \cite{aiello2018decision}} & {\small M6} & {\small %
Fuse contextual information, mathematical} & {\small P2} & {\small O1} & 
{\small Public buildings} \\ 
&  & {\small formulation and experts' knowledge} &  &  &  \\ 
{\small Kiran et al. \cite{R2020113054}} & {\small M7} & {\small Alleviate
the cold start issue} & {\small P1} & {\small O2} & {\small Households} \\ 
{\small Pinto et al. \cite{app10144926}} & {\small M7} & {\small Improve
predictions of consumers'} & {\small P3} & {\small O2} & {\small Public
buildings} \\ 
&  & {\small preferences and recommendations} &  &  &  \\

{\small Rocha et al. \cite{rocha2015improving} } & {\small M8} & {\small %
Decision aggregation of heating and cooling} & {\small -} & {\small O2} & 
{\small Public buildings} \\ 
&  & {\small systems operations with energy demand} &  &  &  \\ 

{\small Anvari et al. \cite{anvari2017multi}} & {\small M8} & {\small %
Multi-agent based optimization} & {\small P1} & {\small O2} & {\small %
Households/} \\ 
&  &  &  &  & {\small public buildings} \\  \hline
\end{tabular}

\end{center}
\end{table}

\section{Critical analysis and discussion} \label{sec4}
First of all, based on the overview of existing energy efficiency recommender systems conducted above and by analyzing the summary in Table~\ref{SummaryRS}, it has been clearly demonstrated that most of the frameworks have been implemented using cloud computing. This is due to its different advantages, among them is its flexible lease and release of computing resources as per the end-user's requirement \cite{ari2019enabling}. However, cloud computing has other issues, such as the privacy preservation problem, which could be occurred whenever the data may outbreak the service provider and the information could be deleted purposely \cite{schaefer2020management}. Additionally, technical issues could also occur due to the fact that servers could be down, and hence, it becomes hard to gain back access to the needed resources/data at the right moment and from anywhere. For instance, non-availability of services could be a result of a denial-of-service attack (DoS) \cite{mahjabin2017survey}.

Furthermore, it is worth noting that most of the recommender systems have been used to polish energy consumption behaviors in households while there are also other frameworks that discussed their utilization in commercial buildings, work spaces and other types of public buildings (e.g. hotels and hospitals). While for the objectives of the recommender systems, almost the same attention has been given for developing both action recommendations and strategy recommendations in existing energy saving recommendation systems frameworks.

On the other hand, recommender systems for energy efficiency face other difficulties and issues, which need to be overcome while developing reliable solutions. In this section, we focus on introducing them along with discussing the commercialization potential of energy saving recommender systems and related issues, i.e. identifying the main market barriers and market drivers \cite{himeur2020marketability}.

\subsection{Limitations and difficulties}
\begin{figure*}[!t]
\centering
\includegraphics[width=\columnwidth]{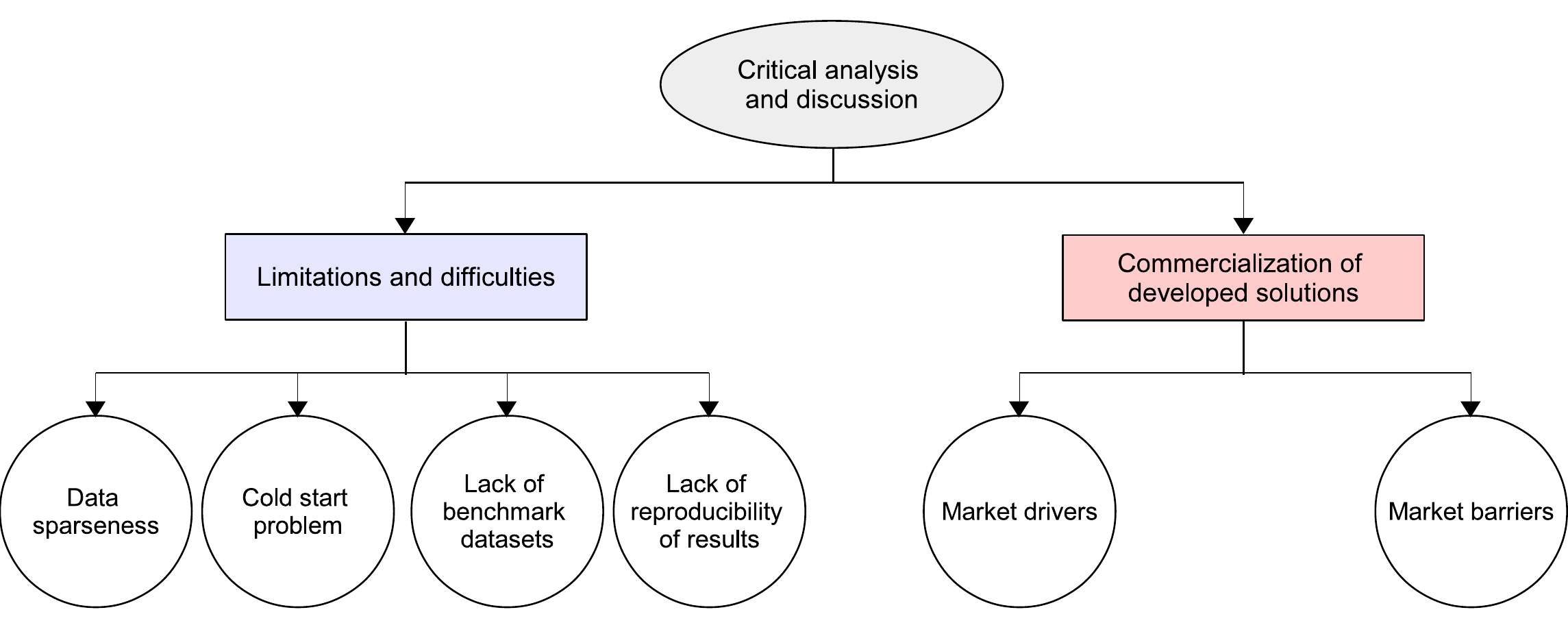}
\caption{A summary of limitations and difficulties of energy saving recommender systems.}
\label{critical_analysis} 
\end{figure*}

Although there has been a significant progress in developing recommender systems as discussed above, various issues that hinder the establishment of effective recommendation engines still exist. The most critical problems and difficulties that exercise a negative impact can be summarized as portrayed in Fig.~\ref{critical_analysis}. Explicitly, it outlines the main difficulties and commercialization issues discussed in the context of energy saving recommender systems.

\subsubsection{Data sparseness}
Recommender systems are mainly based on the analysis of historic consumer data, which usually comprise few customer demographics and mainly customer ratings for items (or actions). Because a consumer may only rate a small number of the actions that are available on the recommendation platform, this leads to a sparseness on the ratings for some actions or users \cite{natarajan2020resolving,jain2020efficient}. Explicitly, this results in producing unreliable recommendations, which in turn could reduce consumer satisfaction \cite{zhang2020alleviating}.

\subsubsection{Cold start problem}
This issue refers to the lack of data (i.e. ratings) for new consumers or new actions. The problem is more acute in the start of the recommendations generation process, when the platform still has a few consumers and limited information about their preferences \cite{son2016dealing}. Explicitly stated user preferences can be used to enhance customer profile, and user similarity metrics can be used to assign newcomers to existing customer clusters with similar preferences. Another problem occurs in systems that refresh their catalog of items or actions. When a new action is added as an option to the system, it does not have any ratings yet, so collaborative methods cannot recommend them. Therefore, this leads to the fact that an action cannot be recommended easily, and hence, less likely to be noticed by consumers \cite{lika2014facing,liu2014promoting}. Probabilistic techniques that recommend new items with lower probabilities and rule-based recommenders can be used in place, to tackle the cold start problem.

\subsubsection{Lack of benchmark datasets}
Due to the large variability in the type of buildings and the respective objectives for energy savings, when developing a recommender system for this purpose, it is of utmost importance to use energy consumption datasets form different environments (e.g. households, commercial buildings and industrial areas) to evaluate the developed systems and efficiently generate personalized recommendations. In this regard, datasets are utilized as benchmarks for developing new recommender models and comparing them to existing systems under the same conditions. Therefore, datasets play a major role in the creation of successful recommendation systems, and most of the effective and robust recommender systems are those built upon large-scale datasets including big amounts of consumers' data \cite{verbert2011dataset}. However, for the energy sector, another important issue that impedes the generation of efficient recommendations is the absence of appropriate datasets along with the difficulties encountered to collect them. Moreover, the lack of open-access repositories containing existing datasets make the recommender systems comparison very difficult, or even impossible \cite{drachsler2010issues, ccano2015characterization}.

\subsubsection{Lack of reproducibility of results}
As presented in Section \ref{sec3.2}, various recommender systems have been developed and utilized for generating energy saving recommendations. However, the comparison of their efficacy is a daunting task since the assessment results could hardly be reproducible due to the absence of toolkits that support such tasks \cite{isinkaye2015recommendation}. Explicitly, the actual issues that hinder reproducing recommendation systems results had put recommender system community in a problematic situation \cite{dacrema2019troubling, ekstrand2011rethinking}. Researchers and developers (who require effective recommender algorithms and baselines against which to compare novel approaches) usually obtain very limited guidance in existing research and development sources \cite{HimeurENB2020}. In this line, to alleviate these issues and enable reproducibility, recommendation systems community needs to (i) review other research topics and inspire from them; (ii) establish a common definition of reproducibility; (iii) capture and comprehend the decisive factors that impact reproducibility; (iv) perform more extensive experiments; (v) promote developing and using recommender frameworks; and (vi) launch experimental platforms that includes recent state-of-the-art algorithms and datasets and establish best-practice guidelines for recommendation system research, which will also ensure a fair comparison among systems \cite{beel2016towards,ie2019recsim}.

\subsection{Commercialization of developed solutions}
\subsubsection{Market drivers}

Energy conservation and energy security have been big concerns in recent years. Energy deficiency does not only affect the economy, culture and development of the world, but also results in global warming. This is why recommender systems play an imperative role in transforming end-user behavior towards improved energy efficiency. Therefore, it is of fundamental importance to investigate the main market drivers that will force the incorporation of energy-efficient and cost-effective technology, and to perform activities that help in controlling the decision-making process for energy efficient buildings, thus, providing an incentive for energy conservation \cite{trianni2013drivers}. An important driver is the EU energy and climate package \cite{climateenergy2030} for the year 2030 that includes the goals of 40\% reduction of greenhouse gas emissions (compared to 1990 levels), and 27\% increase of renewable energy sources (RES) in the EU-27 energy mix (today 6.5\%) and 27\% mitigation in the primary energy consumed (saving 13\% compared to 2006 scales). These goals have led the inception of many energy efficiency research initiatives, some of which incorporate recommender systems \cite{himeur2020data}. 

A global driver, that is derived from the same incentives, is the strategic decision, made by the top management of several firms, to comply with the new energy saving plans. Such decisions include energy reductions within the firm itself, optimizations of the products' and/or services' production pipeline, and efforts to reduce the overall carbon footprint. The TARHSEED initiative, is a bright example: the governmental initiative's aim to raise awareness on energy saving activities and reduce unnecessary energy usage in the country as a whole \cite{kaabi2012conservation, tarhseed_qatar_2020}, has been boosted by several advertising initiatives, standards and competitions, which motivated both individuals and corporations to optimize their energy efficiency.

\subsubsection{Market barriers}

The technology readiness level (TRL), the compatibility (i.e. how feasible it is to integrate it with existing systems), and the business model behind a technological solution are crucial factors in its ability to change its target market. The same factors hold for energy saving recommender systems and affect their impact to the energy market landscape. The global energy landscape is changing, driven by the need to reduce emissions and increase the security of supply while increasing the intermittent renewable energy in the energy mix. 
In this new landscape, the increasing power consumption requires maintaining the power grid reliability, regulating electricity flow with less mismatching between electricity generation and demand and reducing the energy footprint. The optimization of scheduling, the improvement of energy quality and assets efficiency, the integration of dynamic pricing and the incorporation of more renewable electricity sources are among the continuous challenges of the traditional energy grid. 

For a recommender system to penetrate the market and establish its presence, there is a number of market barriers to consider. First, technical barriers present a major milestone in adapting recommender systems in mainstream products and services. Namely, there is a discrepancy between the different evaluation standards of recommender systems, i.e. there is no unified standard for objective benchmarking. Moreover, the lack of comprehensive datasets impedes the progress in creating powerfully dynamic recommenders.

From a legal standpoint, several market barriers exist, which in the context of energy efficiency, comprise a number of regional standards and regulations for residential and industrial energy consumption. For example, in Greece, the Transposition of the European Directive 2009/28/EC was established in 2010. In order to acquire a new permit to build in 2011, it is either appropriate to obtain an annual solar percentage of 60\% for the development of hygienic hot water from solar thermal systems or to explain technological challenges in the event of non-compliance. This, in turn, puts additional obstacles towards commercializing recommender systems as they have to comply with many standards to obtain international recognition.

\section{Current challenges and future orientations} \label{sec5}
\begin{figure*}[!t]
\centering
\includegraphics[width=\columnwidth]{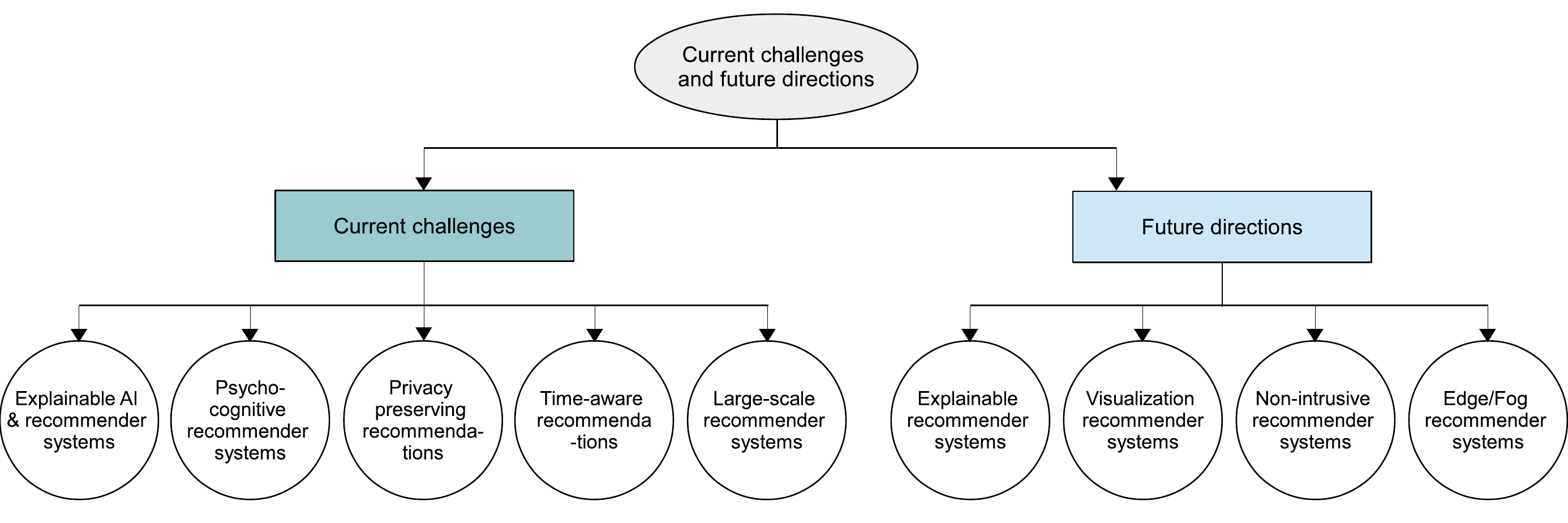}
\caption{Currant challenges and future orientations.}
\label{current_future} 
\end{figure*}

The focus in this section is to provide insights on where the actual energy recommender systems research is heading to as well as the related challenges attracting considerable R\&D in the foreseeable and far future. Specifically, Fig.~\ref{current_future} summarizes the current challenges and future orientations of energy efficiency recommender systems, which have been addressed in this framework.

\subsection{Current challenges}

\subsubsection{Explainable AI \& recommender systems}
\begin{figure*}[!ht]
\centering
\includegraphics[width=\columnwidth]{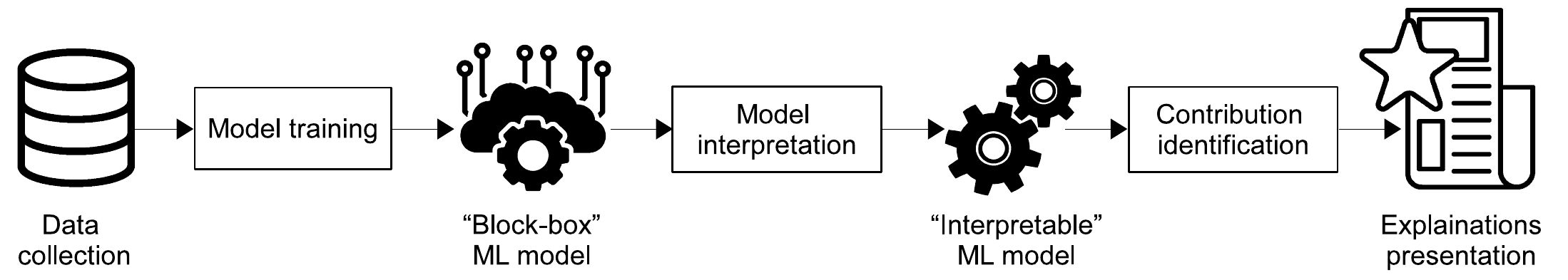}
\caption{The transition to explainable AI.}
\label{fig:transition-of-recommendation-systems}
\end{figure*}

The recent success of machine learning (ML) techniques in solving daily prediction or classification tasks has dramatically influenced the number of applications that adopt ML models, using them as black boxes that makes them difficult to understand by the end-users \cite{gunning2017explainable}. The ability of an ML model to \enquote{explain itself and its actions} to the users is considered to be an emerging important factor for the current modern AI applications transitioning to modern explainable AI models \cite{arrieta2020explainable,WU2021165}, as depicted in Fig.~\ref{fig:transition-of-recommendation-systems}.

In general, there are five key concepts that can describe each recommendation task and are referred to as ``the five \textit{Ws} of recommendations'': $W$hen, $W$here, $W$ho, $W$hat and $W$hy \cite{zhang2020explainable}. The five \textit{Ws} usually represent the time-driven recommendations (\textit{when}), location-driven recommendations (\textit{where}), their social component (\textit{who}), application-driven recommendations (\textit{what}), and their explanations (\textit{why}), respectively.

Following the emergence of explainable AI, the goal of \enquote{Explainable Recommendation Systems} is to offer helpful suggestions to consumers, accompanied by explanations that generally relate to the reasons for providing these recommendations or the advantages of selecting the suggested alternatives \cite{SAGI2020124,fernandez2020random}. The key contribution of these reasons is that they will boost the system's persuasiveness, customer comprehension and happiness and offer an instant reward to the user.

Latest work on explanation-driven recommendations is centered around two core aspects: 1) the type of explanations generated (e.g. textual, visual, etc.); and 2) the employed algorithm or model to generate the explanation, which can be loosely classified as matrix factorization, subject modeling, graph-driven, deep learning, knowledge-graph, interaction rules, and post-hoc models, etc. \cite{zhang2020explainable}. Explainable recommendations, classified by the nature of explanation produced, are enumerated as follows:

\begin{itemize}
\item \textbf{User-based and item-based explanations}: This is a traditional type of explanation based on user's feedback and is expressed as a statement of similarity among the system's different users (in the case of user-based) or items (in the case of item-based recommendations).
\item \textbf{Content-based explanation}: This type is solely based on the item's feature space (e.g. for book recommendations the book type, the writer, etc.).
\item \textbf{Textual explanations}: In the textual explanations, the recommendations include explanation sentences that may be based on other users' reviews or natural language processing techniques.
\item \textbf{Visual explanations}: Visual explanations utilize item images for explainable recommendations indicating the part of the image that matches the item images that the user might be interested in.
\item \textbf{Social explanations}: These explanations refer to items that user's \enquote{friends} in the social networks or in a specific community also prefer.
\item \textbf{Hybrid explanations}: Hybrid explanations refer to combinations of one or more of the previous types of explanations.
\end{itemize}

Focusing on energy saving recommender systems, currently we identify the lack of recommendation systems in the area of energy sustainability, which adopt the recent trends of explainable AI. 
Recent surveys like \cite{zhang2018explainable} try to overview existing methods to set the current research issues related to the explainable recommendations. For instance, \cite{gao2019explainable} proposes a deep explicit attentive multi-view learning architecture for modeling multi-level characteristics of explanations, or the framework in \cite{balog2019transparent} that examines another scheme to create a set-based recommendation platform to generate textual and transparent explanations of film recommendations. Aiming at developing a knowledge-based scheme to create explainable item recommendations, a technique to leverage external knowledge is proposed in \cite{catherine2017explainable}, which is based on adopting knowledge graphs when information from content and product/item reviews is unavailable for generating explanations. Interpretable models, are based on transparent processes to decide the recommendation lists, hence, it is easier for generating appropriate explicit feature-level explanations to justify the reasons behind the recommender's suggestion for particular items \cite{zhang2014explicit}. Following the same scenario of graph-based models, He et al. in \cite{he2015trirank} introduce a technique that could rank the vertices of a tripartite graph and furnish explanations for the top-ranked, aspects-target, and user-recommended item triplets. By contrast, in the field of energy saving and recommendations for energy-related behavior, there is limited literature that elaborates on the rules of producing a particular recommendation. Authors in \cite{grimaldo2019user} propose a user-centric and visual analytics approach for developing an interactive and explainable forecasting and analysis of electric power demand in prosumer settings. Moreover, it has been advocated that this would be endorsed by behavioral analysis to enable the treatment of possible relationships between energy usage footprints and the interaction of prosumers with energy analysis tools, including customer portals and recommendation systems. 

\subsubsection{Psycho-cognitive recommender systems}
Most of the actual tailored energy recommender systems are, by and large, limited in terms of effectiveness \cite{sardianos2020data}. To improve their performance, the incorporation of cognitive and behavioral knowledge, is essential, where tailored recommendation frameworks can be friendlier and more human-centric \cite{shi2015towards,zhao2014context}. Thus, this results in eventually enhancing users' experience and loyalty and increase their satisfaction. To that end, more effort should be established in this direction aiming at developing psycho-cognitive method recommendation systems that generate personalized energy saving actions and advice based on consumers' preferences, emotional states and centers of interest \cite{shafto2016human,kopeinik2016improving}.

\begin{figure}[!t]
\centering
\includegraphics[width=\columnwidth]{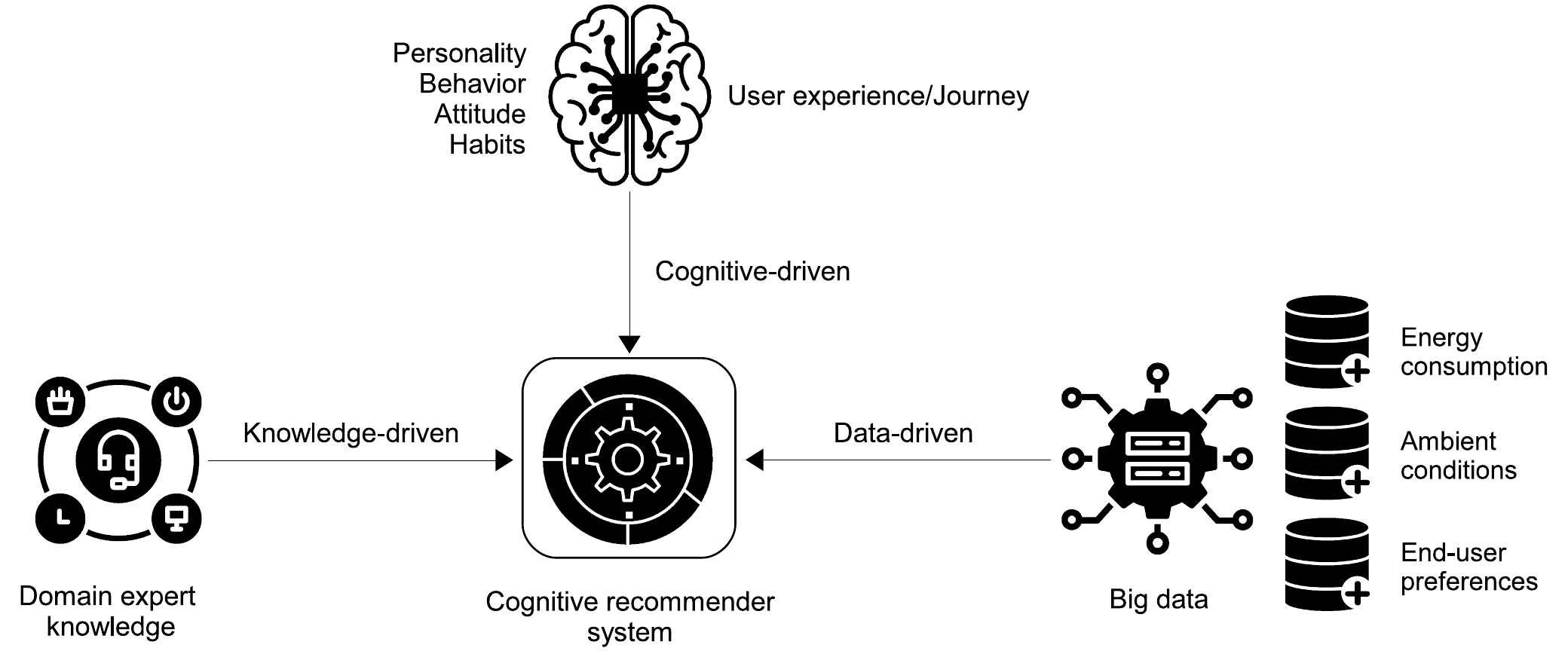}
\caption{Typical representation of a cognitive recommender system for energy saving.}
\label{cog-rec} 
\end{figure}

Accordingly, psycho-cognitive recommender systems are new intelligent recommender systems that help in (i) comprehending end-users' preferences; (ii) detecting changes in end-users' habits and attitudes through time; (iii) predicting end-users' unknown choices and behavioral change; and (iv) investigating adaptive techniques for generating intelligent recommendations within a changing environment \cite{hamlabadi2017framework,kopeinik2017applying}. All these tasks mixed together could improve end-users' behaviors and increase their awareness towards a more sustainable and energy-efficient usage \cite{aguilar2017general,beheshti2020towards}. A typical representation of psycho-cognitive recommender system is illustrated in Fig.~\ref{cog-rec}, in which three important mechanisms called knowledge-driven, cognition-driven and data-driven are used to develop a psycho-cognitive recommendation framework.

The same category also includes recommender systems that build on persuasiveness in order to maximize acceptance \cite{bothos2016recommender}. In some cases, the messages sent to the user are also personalized to match user's preferences and values \cite{sardianos2020emergence}. Active learning is a key component in such approaches because it allows the system to continuously adapt to the changing user needs and demands \cite{sanchez2020persuasion}.

\subsubsection{Privacy preserving recommendations}
The preservation of user privacy is a key requirement in several recommender systems, especially in online social communities. Several techniques have been proposed in the past, ranging from k-anonymity, to differential privacy and homomorphic encryption.
This kind of frameworks could be split into three main groups: (i) perturbation-based techniques that introduce noise to the existing data \cite{puglisi2015content}, without affecting the final recommendation result; (ii) encryption-based schemes that transform the original information within the main recommendation technique (e.g. within the matrix factorization component \cite{tang2017privacy}); and (iii) techniques that develop novel matrix factorization algorithms under local differential privacy (LDP) \cite{shin2018privacy}.
In the case of content recommender systems, group-based approaches \cite{li2017efficient} implement the principles of k-anonymity in order to maintain recommendation efficiency without affecting user privacy. Similar challenges are set for navigation solutions that rely on crowd-sourced data collection \cite{tseng2016privacy}.

Another essential challenge in privacy-preservation is servers-related, and tackles their features, especially when they are unreliable (untrusted) or comprise security weaknesses (vulnerabilities), and thereby collecting consumers' feedback may result in cyber liability owing to data leakage \cite{wang2016trust}. 
Early works in the privacy of consumer data in electric load monitoring applications \cite{mclaughlin2011protecting} mostly focus on non-intrusive monitoring techniques to combat potential invasions of privacy \cite{himeur2020smart}. Later works minimize the amount of collected reference data through sampling. For example, in \cite{englert2015enhancing} authors remove redundant energy traces, which do not contribute new knowledge to the recommender system.

\begin{figure*}[!t]
\centering
\includegraphics[width=\columnwidth]{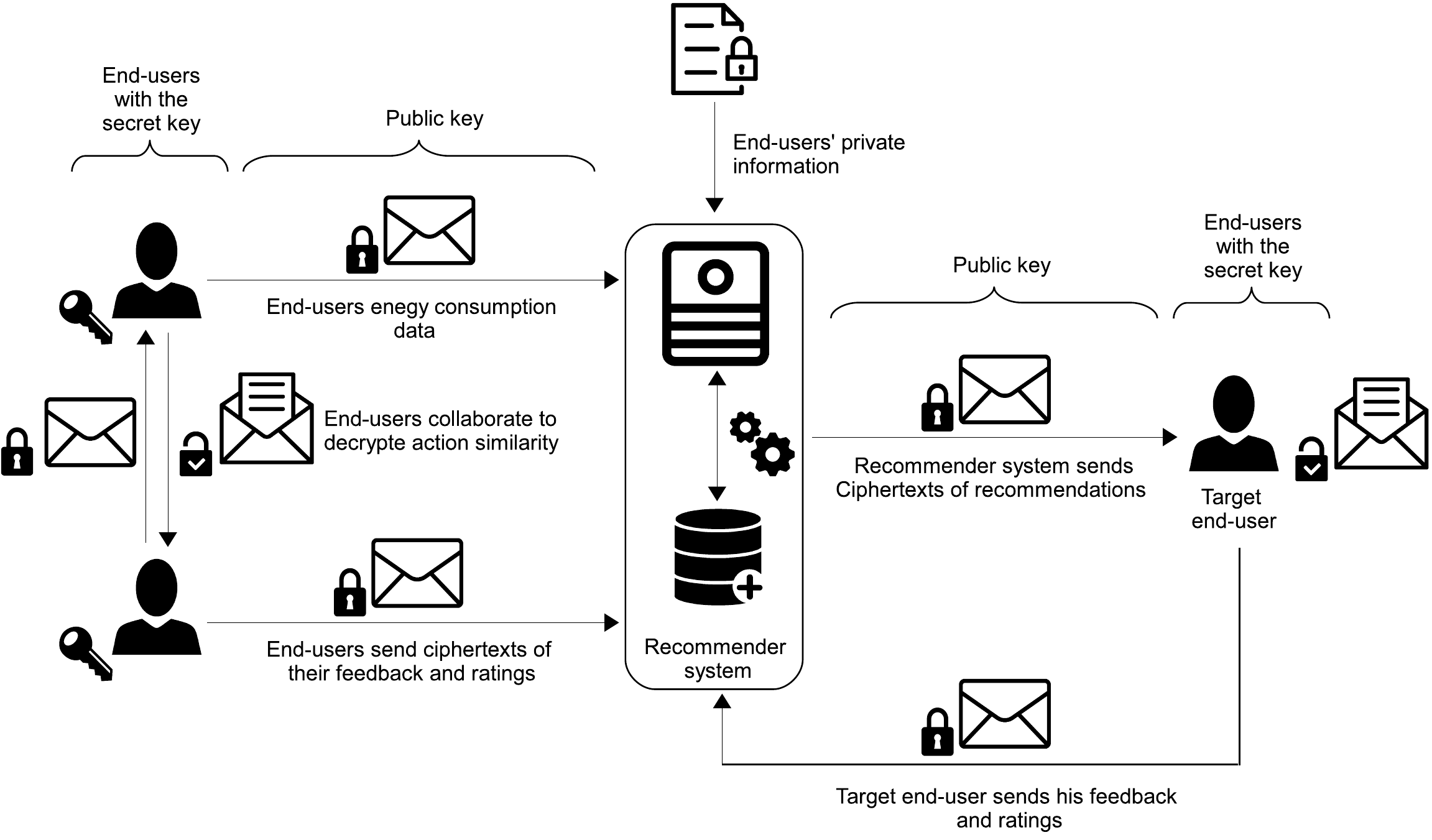}
\caption{A typical privacy-preserving recommender system based on encryption and collaborative filtering.}
\label{priv-rec} 
\end{figure*}

To the same direction end, privacy-preserving recommendation approaches aim at preserving consumers' privacy through hiding their rating feedback from servers and/or other consumers \cite{badsha2017privacy,jiang2019towards}. Fig.~\ref{priv-rec} presents a typical representation of an energy saving recommender systems. It illustrates what are the sensitive information that need to be encrypted before submitting them to the recommender system's server, such as energy consumption data (provides the intruder with information about the presence/absence of the end-user in his household), end-users' feedback and ratings and private information (personal data, location, number of end-users, etc.) \cite{badsha2016practical,xu2018privacy}. After storing and processing collected data, the recommender system encrypts the generated recommendations before sending them to the targeted end-user. Moreover, end-users could collaborate with each other to compute action similarity using their private keys. On the other hand, with the arrival of the blockchain technology, new opportunities have been opened up to develop a novel generation of recommender systems that can overcome the privacy-preservation issues and protect consumers' data \cite{begum2019towards,pu2020efficient}. For example, Bosri et al. propose Private-Rec, which is privacy-preserving recommender system using AI and blockchain \cite{bosri2020integrating}. Explicitly, blockchain has been used to provide the end-user with a secure mechanism using the distributed attribute where data could be exchanged with the required permission. While in \cite{casino2019efficient}, blockchain is deployed as the backbone of a decentralized recommender system, in which a secure architecture has been introduced using decentralized locality sensitive-hashing classification along with recommendation generation.

\subsubsection{Time-aware recommendations}

The concepts of time-aware and concept-aware recommendations have been widely discussed in the recommender systems' community. From the early works on movie recommendations \cite{gantner2010factorization}, to the more recent works on time-aware point-of-interest recommendations \cite{yuan2013time, zhang2015ticrec}, several frameworks for modeling, computing and presenting time-aware recommendations have been proposed in the general domain \cite{stefanidis2013framework, campos2014time}.

Recommender systems for the energy sector differ highly from those used in other research topics. Explicitly, most existing models concentrate on recommending energy saving actions that fit consumers' preferences while putting a slight importance on the temporal information and its influence on the recommendations \cite{linda2020effective}. In this respect, we assert that further attention should be paid for time-aware recommendations in energy saving applications in buildings to push them into the foreground \cite{wang2018personalized,qian2019ears}. This kind of recommendations is more appropriate to emergency situations, e.g. the case of the Coronavirus COVID-19 pandemic, where real-time and time-aware recommendations should be provided according to the current situation. Explicitly, due to the mass restrictions imposed on people's movement, the rise of teleworking and online learning has led to an increase of energy consumption in domestic buildings \cite{beheshti2020towards,nilashi2020intelligent}. On the other side, with the widespread use of ML tools, using deep learning models would be a promising approach to develop recommender systems that encompass contextual information into neural collaborative filtering recommender frameworks \cite{lamche2015context,unger2020context}.

\subsubsection{Large-scale recommender systems}
As we are in the big data era, modern energy saving recommendation systems face tremendously increasing data volume and complexity due to the use of a massive number of connected devices. Traditional computation algorithms and experiences on small datasets may not be efficient today. Therefore, developing robust recommender systems that is capable of processing large-scale data, is becoming a challenging endeavour for several applications. Authors in \cite{eirinaki2018recommender} provide an interesting survey on the challenges and solutions for recommender systems for large-scale social networks. Big data, variety, and volume are the three major challenges for recommender systems in large social networks, which bring state-of-the-art collaborative filtering algorithms to their limits. Additionally, the large volatility of social network data (e.g. new users and items added or removed on a daily basis) has raised the interest for new evaluation metrics, that promote recent \cite{sanchez2018time} and diverse \cite{kunaver2017diversity} entries (e.g. diversity, freshness, serendipity, etc.) and tackle the cold start problem.

From the ML and deep learning perspective \cite{batmaz2019review}, graph convolutional methods are gaining the researchers' interest \cite{ying2018graph}, since they can summarize the graph structure of social networks and combine it with the lateral information that may be hidden in the items or in the relations among them. Compact latent factor models that combine content with ratings \cite{liu2016large} prove to be more efficient than simple collaborative filtering algorithms in tackling the cold start problem. In order to balance the exploration-exploitation dilemma (exploit interesting items while exploring new items) and continuously capture user feedback without relying on item context, multi-armed bandit approaches \cite{barraza2020introduction, zhou2017large, canamares2019multi, sanz2019simple} have been recently proposed.

In order to solve the technical issues that may arise from the scalability of recommender systems in large datasets, several parallel and distributed algorithms have been proposed, which either rely on the splitting of the dataset, using social or other information \cite{sardianos2017scaling, bathla2020scalable} and its parallel processing, or on the refactoring of existing algorithms in order to take advantage of the use of graphical processing units (GPUs) \cite{li2019efficient, li2017msgd, sardianos2018survey}.
The issue of big data handling has also been studied in the domain of energy efficient recommender systems for recommending energy plans \cite{zhang2018big}, providing actionable recommendations \cite{sardianos2020data} or improving comfort and energy efficiency in tandem \cite{kar2019revicee}. The solutions discussed so far in the pertinent literature focus on data sampling or compression.

\subsection{Future orientations}
This section highlights the most promising research directions that will have a significant impact on improving the effectiveness of energy saving recommender systems in the near future.

\subsubsection{Explainable recommender systems}
As described before, increasing the user's trust and transparency to the system is an important concept in modern ML models and also a tool for maximizing the recommendations' acceptance in modern recommender systems. Thus, in the field of recommender systems for energy saving, the system has to accompany every recommendation for an energy saving action with: 
\begin {enumerate} 
\item an explanation of \textbf{why} this particular advice is suggested
\item a statistical fact on \textbf{what} are the benefits following the recommended action
\end {enumerate}

\begin{figure*}[!ht]
\centering
\includegraphics[scale=0.64]{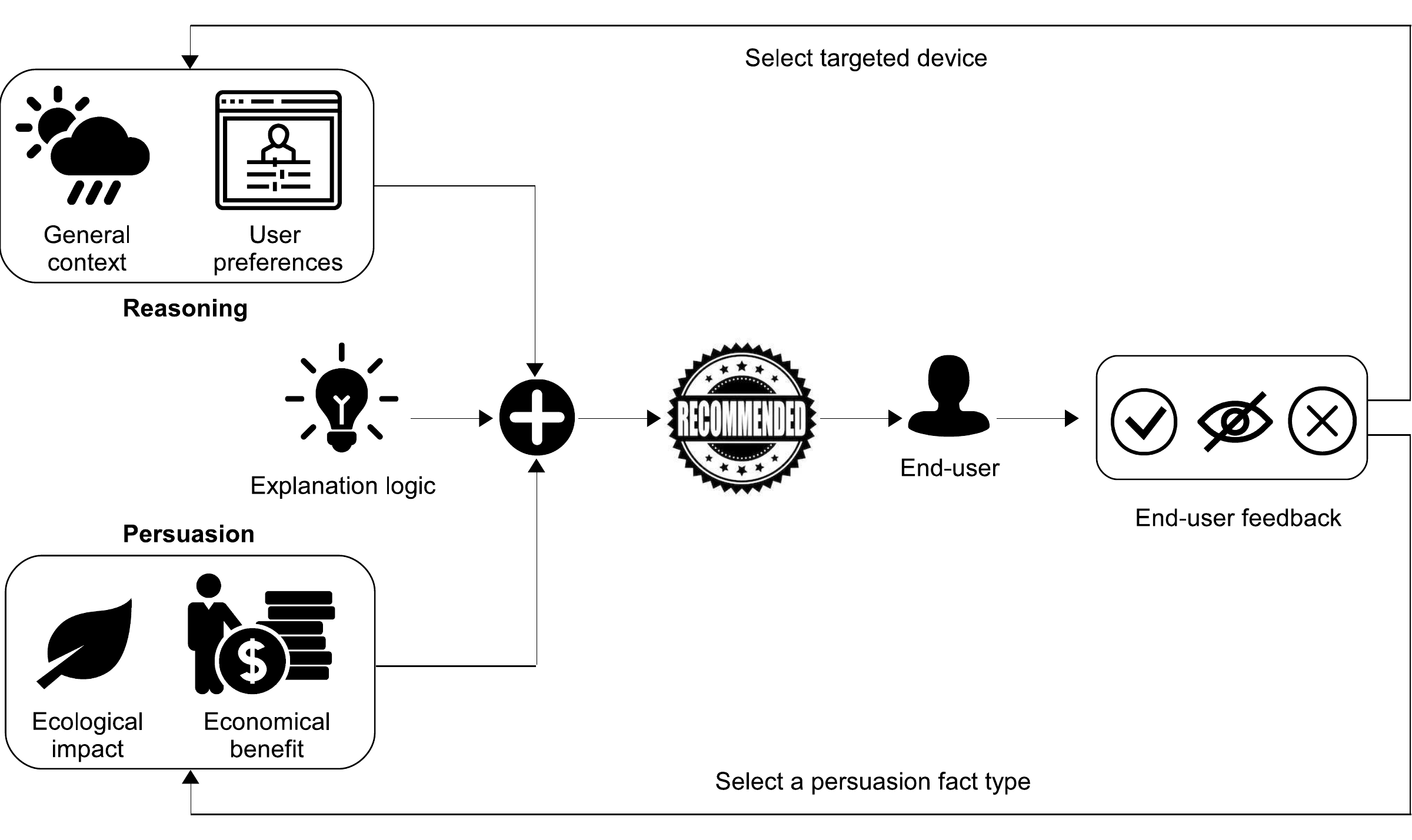}
\caption{The flow of explainable recommendations generated in the (EM)\textsuperscript{3} framework.}
\label{fig:explRecommendations-methodology}
\end{figure*}

Hence, we have introduced the (EM)\textsuperscript{3} explainable recommender system for energy efficiency\footnote{(EM)$^3$: Consumer Engagement Towards Energy Saving Behavior by means of Exploiting Micro Moments and Mobile Recommendation Systems (\url{http://em3.qu.edu.qa/})}, and described the two most essential characteristics that recommendation explainability is based on, as depicted in Fig.~\ref{fig:explRecommendations-methodology} \cite{SARDIANOS2020394,sardianos2020smart}:
\begin {enumerate}
\item \textbf{Reasoning:} This feature explores the context of global recommendations and seeks to provide thorough justification of why each recommendation has been made. It may be metadata of the \textit{end-user context} (e.g. the end-user is not present in the room), about the \textit{device consumption} (e.g. it is turned on for a long time) or about the \textit{external circumstances} that cause the appliance to be switched off.

\item \textbf{Persuasion:} This aspect draws on end-users' expectations, motivations and long-term values, and adopts their ratings to pick the most suitable and relevant explanation about each recommended intervention.
\end{enumerate}

Based on this strategy, and adopting a hybrid type of explanation in our approach, we enhance a persuasion fact along with the textual explanation in the recommendations' body. Using this approach, we try to provide the actual benefit the recommended action will achieve for the end-user, in an attempt to persuade him/her to increase his acceptance over the provided recommendations. Initial results of our evaluations show that such an approach can impact users' trust to the system and can bring an increase of 20\% in the acceptance ratio of provided recommendations.

\subsubsection{Visualization recommender systems}
In this technological age, it is not a secret that humans are attracted to imagery type of media, much more than the ones of textual nature. This can be witnessed by how millennials are exploiting technology nowadays. With this in mind, it can be argued that in order to have a better recommendation dialogue with end-users, aiding such recommendations with visualized charts and evidence can significantly aid in making them persuasive. By stating this, it is by no means indicating that the textual recommendations, i.e. explanations provided by the recommenders, should be discarded, but rather they are complementary to one another. Visualization and textual interpretations hand-in-hand can be integrated fully to structure suggestions given by the recommender systems, which are deemed to influence behavioral change. All the following discussed frameworks have used visualizations one way or the other to influence behavioral change in their system.

\begin{figure*}[!htb]
\centering
\includegraphics[width=\columnwidth]{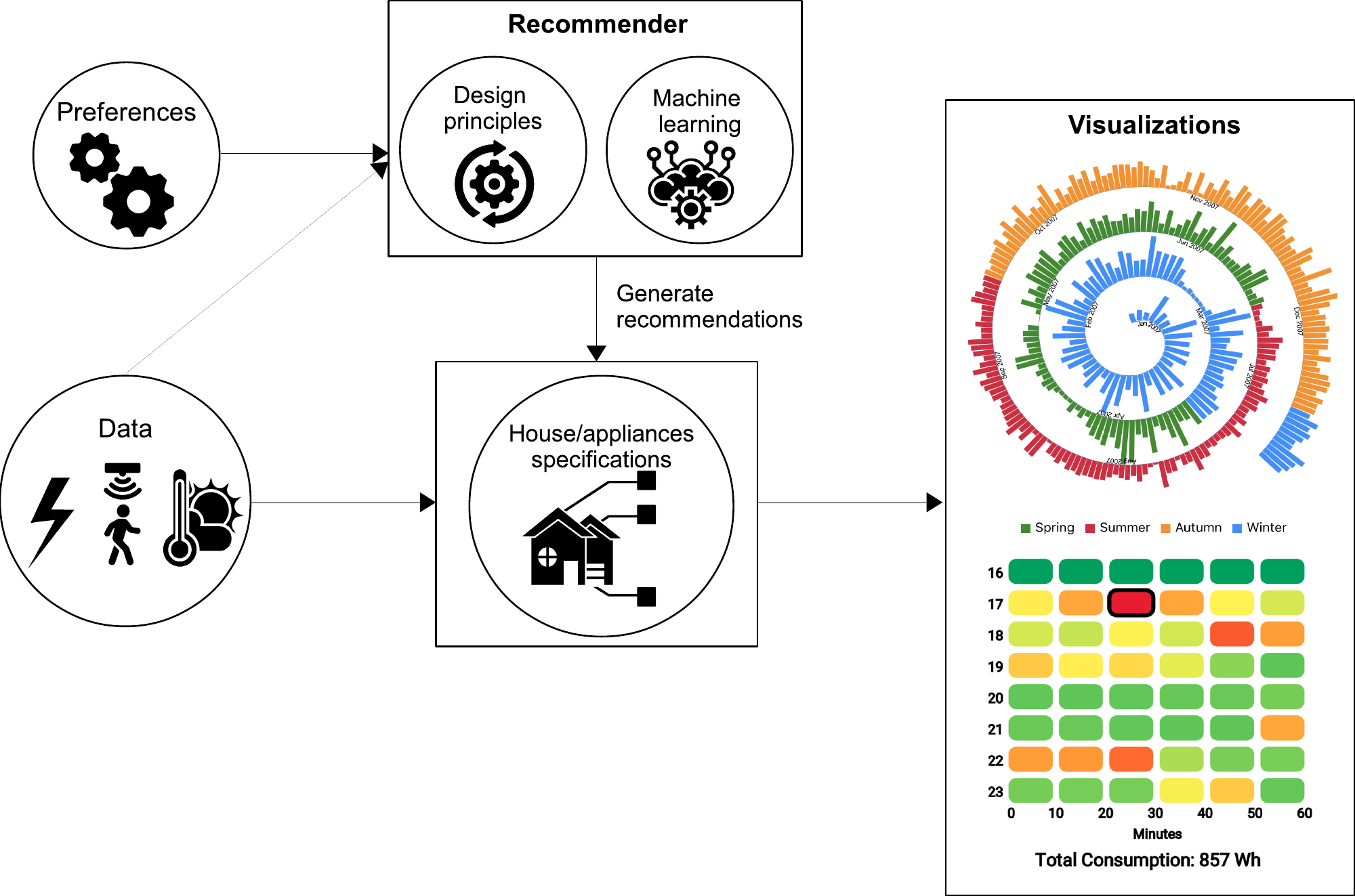}
\caption{Example of the visual recommender system proposed in the (EM)\textsuperscript{3} framework.}
\label{vis-rec} 
\end{figure*}

A semantic smart home system for energy efficiency, namely SESAME, is proposed in \cite{Fensel2013} to provide daily and monthly overlays of energy consumption data, CO$_{2}$ footprint and financial impact. Also, the user interface (UI) to control the appliances and create certain rules is also shown. On the other hand, Fernández et al. \cite{RodriguezFernandez2016} showcase a heatmap exhibiting, in hourly fashion, the usage of air conditioning facilities for a whole week. Similarly, friendly UI is also provided to allow users more control over given services through smart devices. On the other hand, enCOMPASS framework push visualizations aided by context-aware collaborative recommender systems on mobile platforms to provide energy-efficient recommendations from socio-technical point-of-view \cite{Fraternali2017,Fraternali2018}. Moreover, the framework also created two games on smartphones to teach children about the importance of rationalizing the consumption of both water and energy. The former being taught through ``Drop! The Question'' application, which is developed with SmartH2O framework, and the latter through ``Funergy'' application \cite{Albertarelli2017}. Entropy, another framework, provides conventional time-series visualization for sensor data streaming in a desktop application to aid the recommender system \cite{Fotopoulou2017}, while both Bernard and HEMS-IoT create a mobile application for that purpose \cite{Zorrilla2019,Machorro-Cano2020}. CASER framework produces both web and mobile applications for data visualization but showcase variant visualization including time-series and heatmap charts on both household and substation levels (multiple households) \cite{Sitoula2019}. Lastly, (EM)\textsuperscript{3} creates two distinct applications, where the first application on iOS showcases data visualization in recommendations \cite{Alsalemi2020} and the latter on both Android and iOS, developed on React Native, studies the effect of different charts on end-users understanding \cite{Al-Kababji2020}. Fig.~\ref{vis-rec} depicts the flowchart of the visualization recommender system developed in the (EM)\textsuperscript{3}.

From the previous illustrated work, three important prospects can be further investigated when integrating the data visualization pillar with the recommendations. Firstly, further studies can be established to understand the impact different visualizations have on end-users as in \cite{al2020interactive}. Not only that, but also create novel data visualizations specifically for energy consumption data, which are deemed simple for end-users from different backgrounds. Secondly, using the visualization graph hand-in-hand when the recommendation is suggested. In other words, the recommender system refers to the visualization and demonstrates the anomalous consumption through such visualizations. Thirdly, with the recent pandemic humanity is facing (i.e. COVID-19), people are working from home for social distancing, and thus, the energy consumption has increased in the domestic sector \cite{BBC2020}. Therefore, it would be more important than ever to generate personalized and timely recommendations.

\subsubsection{Non-intrusive recommender systems}
Non-intrusive load monitoring (NILM) has been deployed in different energy saving projects instead of submetering for detecting appliance-level consumption data and other related information, e.g. when exactly a specific appliance is turning on/off without the need to install further submeters \cite{Himeur2020IJIS-NILM,siddiqui:hal-02954362}. In this line, collecting energy consumption feedback using NILM has turned around good performance at low/no cost. On one hand, developing efficient energy saving recommender systems relies on an accurate analysis of energy consumption data, especially at the appliance-level, and on the detection of abnormal energy usage. Tailored recommendations are then produced following the feedback of the anomaly detection stage.
Thus, the development of non-intrusive recommendation systems using submetering is a promising solution to be investigated, which is not scalable because it needs to analyze individual appliance consumption traces \cite{himeur2020detection}. On the other hand, for energy data analysis and visualization, NILM has been considered as a scalable and practical alternative to submetering. However, the use of NILM in recommender systems has not been discussed before since the aim of NILM is to provide appliance-level energy footprints. In this regard, it is of significant interest to assess the signal fidelity of devices' fingerprints generated by existing NILM algorithms to develop effective non-intrusive recommender systems. This could figure out end-users' preferences and related information as well \cite{himeur2020anomaly}. Consequently, by using NILM instead of submetering, the development cost of recommender systems will significantly be reduced \cite{Luo2017,Himeur2020IJIS-AD}.

\subsubsection{Edge/Fog recommender systems}
Energy recommender systems have become an essential solution for energy efficiency in buildings. While a large number of existing frameworks are focused on using cloud-to-edge architectures, in which recommended energy saving actions are transmitted to the edge device (e.g. consumer's smartphone) after completing the computing task in the cloud server \cite{gong2020edgerec}. Although these architectures allow to achieve a good efficiency, they are prone to serious noticeable delays in the system's feedback and user's reaction because of the network bandwidth and latency between the cloud and edge \cite{sun2020convergence}. 
By contrast, implementing the recommender algorithms directly on the edge can allow real-time computing and identify consumers' interests/preferences more accurately and thereby increasing their satisfaction and trust of the generated recommendations \cite{su2019edge,himeur2020emergence}. Thus, recently, a great deal of attention is devoted to develop and implement recommender systems on the edge and/or on fog devices, which can tremendously reduce the computational time, minimize the cost of cloud hosting and ensure privacy-preservation \cite{hidasi2018cutting,felfernig2017recommendation,sayed2021endorsing}.

\begin{figure*}[!t]
\centering
\includegraphics[width=\columnwidth]{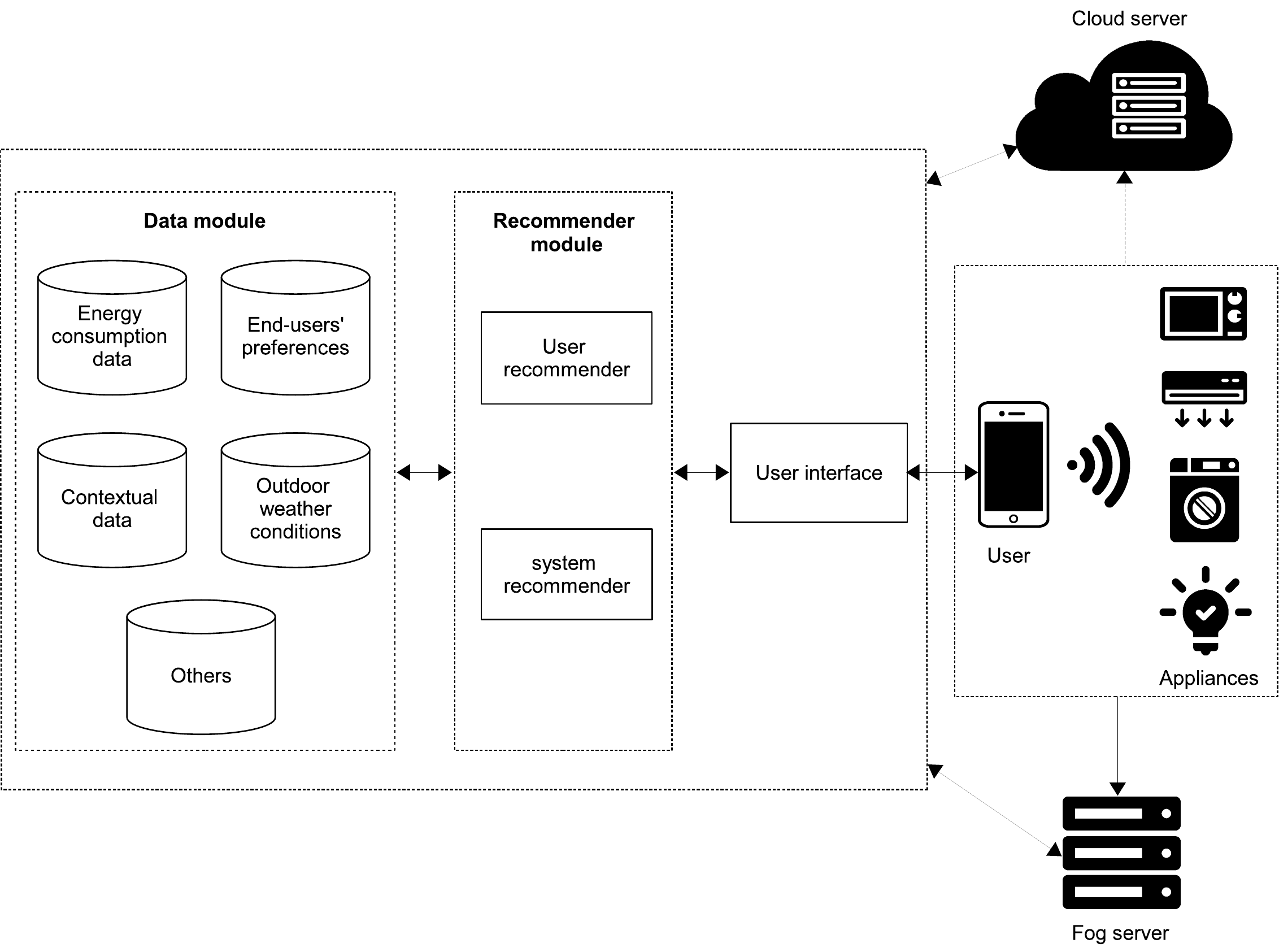}
\caption{Example of a typical representation of a fog-based energy recommender system.}
\label{fog-rec} 
\end{figure*}

To that end, various frameworks have been proposed in different research fields to investigate the applicability of edge-based and fog-based recommender systems \cite{wang2020privacy,zhou2020cost}.
For instance, in \cite{wang2019fog}, aiming to satisfy the new requirements of recommender systems, e.g. the low latency and uninterrupted service, Wang et al. propose a fog-based recommendation framework based on collaborative filtering. It can overcome the problem of information overload in fog computing and help in generating personalized recommendations for improving system performance. In the same manner, a fog-based recommender system which helps to bridge the gap between the cloud and end-devices is proposed in \cite{ibrahim2020fog}. This system has been used to improve the performance of the E-Learning environments. Furthermore, in \cite{jabeen2019iot}, a fog-based recommender system is introduced to provide recommendations regarding the lifestyle, dietary plans and exercises to an ensemble of cardiovascular disease patients. While in \cite{gong2020edgerec}, an edge-based recommender system is proposed to allow (i) capturing real-time end-users' preferences more precisely; and (ii) generating personalized and satisfying recommendations. Fig.~\ref{fog-rec} describes a typical representation of a fog-based recommender system for multiple users, in which a hybrid computing scheme, using cloud and fog servers, is generally adopted to implement the main tasks of the recommendation framework.

\section{Conclusions} \label{sec6}
This article presents a critical review of recommender systems for energy efficiency in buildings. Accordingly, a taxonomy of recommender systems is firstly conducted based on different aspects. Following, a critical analysis is conducted to highlight their strengths and limitations before deriving the current challenges and cutting-edge topics, which can be targeted in the near future to improve their performances.

By and large, energy saving recommendation systems are proving to be a promising solution to promote sustainability and reduce carbon emissions, especially with the widespread deployment of smart-meters, IoT sensors and ML tools. Their evolution is accompanying the evolution of the intelligent Internet systems. The first generation of recommendation frameworks were based on collaborative filtering, case-based, PRM and Rasch-based engines through analyzing only energy-based data. While the second generation relies on the use of information fusion and deep learning models, in which other kinds of data are gathered and transmitted to the recommender engine to be analyzed together with energy consumption footprints. This movement, that the second generation is promoting for, helps in generating more accurate recommendations.

Moving forward, we have also discussed the third generation of recommender systems for energy efficiency, which relies on adding other innovative modules into the recommender engines, i.e. explanations, visualizations and time-aware information processing. The use of edge computing technologies and edge AI is playing a major role in making development real-time recommendation systems a reality. Moreover, this results in improving the quality and acceptance of recommendations and increasing the end-users' satisfaction. In this line, future research will focus on fostering the existing systems and technologies for improving both the quality and applicability of recommendation frameworks. Concurrently, novel directions of research will be furthered to develop a novel generation of highly automated recommender system via the use of (i) NILM strategies instead of conventional smart-metering; (ii) edge computing as an alternative to cloud computing; and (iii) privacy-preservation recommendation systems to increase end-users' trust. 

Finally, it is worth noting that the application of recommender systems in the building energy sector is a very promising field since it does not only recommend energy saving actions but can also be extended to help consumers acquire appliances. In this regards, several factors could impact the choice of the consumer, such as the energy consumption of the appliance, its manufacturer, its price and other specifications. However, this will make the development of a more comprehensive energy efficiency recommender systems more challenging. In the grand scheme, recommender system will remain a strong pillar in the future of artificial intelligence and behavior change.

\section*{Acknowledgements}\label{acknowledgements}
 This paper was made possible by National Priorities Research Program (NPRP) grant No. 10-0130-170288 from the Qatar National Research Fund (a member of Qatar Foundation). The statements made herein are solely the responsibility of the authors.



\begin{thebibliography}{100}

\bibitem{cao2016building}
X.~Cao, X.~Dai, J.~Liu, Building energy-consumption status worldwide and the
  state-of-the-art technologies for zero-energy buildings during the past
  decade, Energy and buildings 128 (2016) 198--213.

\bibitem{Himeur2020AE}
Y.~Himeur, A.~Alsalemi, F.~Bensaali, A.~Amira, Robust event-based non-intrusive
  appliance recognition using multi-scale wavelet packet tree and ensemble
  bagging tree, Applied Energy 267 (2020) 114877.

\bibitem{economidou2020review}
M.~Economidou, V.~Todeschi, P.~Bertoldi, D.~Agostino, P.~Zangheri,
  L.~Castellazzi, {Review of 50 years of EU energy efficiency policies for
  buildings}, Energy and Buildings (2020) 110322.

\bibitem{pylsy2020buildings}
P.~Pylsy, K.~Lylykangas, J.~Kurnitski, Buildings’ energy efficiency measures
  effect on co2 emissions in combined heating, cooling and electricity
  production, Renewable and Sustainable Energy Reviews 134 (2020) 110299.

\bibitem{Himeur2020iscas}
Y.~Himeur, A.~Alsalemi, F.~Bensaali, A.~Amira, Efficient multi-descriptor
  fusion for non-intrusive appliance recognition, in: 2020 IEEE International
  Symposium on Circuits and Systems (ISCAS), IEEE, 2020, pp. 1--5.

\bibitem{refat2020prospect}
K.~H. Refat, R.~N. Sajjad, Prospect of achieving net-zero energy building with
  semi-transparent photovoltaics: A device to system level perspective, Applied
  Energy 279 (2020) 115790.

\bibitem{huang2018uncertainty}
P.~Huang, G.~Huang, Y.~Sun, Uncertainty-based life-cycle analysis of near-zero
  energy buildings for performance improvements, Applied Energy 213 (2018)
  486--498.

\bibitem{lin2020towards}
Y.~Lin, S.~Zhong, W.~Yang, X.~Hao, C.-Q. Li, Towards zero-energy buildings in
  china: A systematic literature review, Journal of Cleaner Production (2020)
  123297.

\bibitem{Himeur2020IJIS-NILM-R}
Y.~{Himeur}, A.~{Elsalemi}, F.~{Bensaali}, A.~Amira, Recent trends of smart
  non-intrusive load monitoring in buildings: A review, open challenges and
  future directions, International Journal of Intelligent Systems (2021) 1--28.

\bibitem{azizi2019making}
Z.~M. Azizi, N.~S.~M. Azizi, N.~Z. Abidin, S.~Mannakkara, Making sense of
  energy-saving behaviour: A theoretical framework on strategies for behaviour
  change intervention, Procedia Computer Science 158 (2019) 725--734.

\bibitem{staddon2016intervening}
S.~C. Staddon, C.~Cycil, M.~Goulden, C.~Leygue, A.~Spence, Intervening to
  change behaviour and save energy in the workplace: A systematic review of
  available evidence, Energy Research \& Social Science 17 (2016) 30--51.

\bibitem{belhadi65deep}
A.~Belhadi, Y.~Djenouri, G.~Srivastava, D.~Djenouri, J.~C.-W. Lin, G.~Fortino,
  Deep learning for pedestrian collective behavior analysis in smart cities: A
  model of group trajectory outlier detection, Information Fusion 65 (2021)
  13--20.

\bibitem{fraternali2017encompass}
P.~Fraternali, S.~Herrera, J.~Novak, M.~Melenhorst, D.~Tzovaras, S.~Krinidis,
  A.~E. Rizzoli, C.~Rottondi, F.~Cellina, encompass—an integrative approach
  to behavioural change for energy saving, in: 2017 Global Internet of Things
  Summit (GIoTS), IEEE, 2017, pp. 1--6.

\bibitem{casals2017serious}
M.~Casals, M.~Gangolells, M.~Macarulla, A.~Fuertes, V.~Vimont, L.~M. Pinho, A
  serious game enhancing social tenants’ behavioral change towards energy
  efficiency. in 2017 global internet of things summit (giots) (2017).

\bibitem{Sardianos2020GreenCom}
C.~{Sardianos}, C.~{Chronis}, I.~{Varlamis}, G.~{Dimitrakopoulos}, Y.~{Himeur},
  A.~{Alsalemi}, F.~{Bensaali}, A.~{Amira}, Real-time personalised energy
  saving recommendations, in: The 16th IEEE International Conference on Green
  Computing and Communications (GreenCom), 2020, pp. 1--6.

\bibitem{hwang2016efficient}
W.-S. Hwang, H.-J. Lee, S.-W. Kim, Y.~Won, M.-s. Lee, Efficient recommendation
  methods using category experts for a large dataset, Information Fusion 28
  (2016) 75--82.

\bibitem{Himeur2020icict}
Y.~Himeur, A.~Alsalemi, F.~Bensaali, A.~Amira, Improving in-home appliance
  identification using fuzzy-neighbors-preserving analysis based
  qr-decomposition, in: International Congress on Information and Communication
  Technology, Springer, 2020, pp. 303--311.

\bibitem{Himeur2020IntelliSys}
Y.~Himeur, A.~Alsalemi, F.~Bensaali, A.~Amira, C.~Sardianos, I.~Varlamis,
  G.~Dimitrakopoulos, On the applicability of 2d local binary patterns for
  identifying electrical appliances in non-intrusive load monitoring, in:
  Proceedings of SAI Intelligent Systems Conference, Springer, 2020, pp.
  188--205.

\bibitem{becchio2018impact}
C.~Becchio, M.~Bertoncini, A.~Boggio, M.~Bottero, S.~P. Corgnati,
  F.~Dell’Anna, The impact of users’ lifestyle in zero-energy and emission
  buildings: An application of cost-benefit analysis, in: International
  Symposium on New Metropolitan Perspectives, Springer, 2018, pp. 123--131.

\bibitem{ashouri2018development}
M.~Ashouri, F.~Haghighat, B.~C. Fung, A.~Lazrak, H.~Yoshino, Development of
  building energy saving advisory: A data mining approach, Energy and Buildings
  172 (2018) 139--151.

\bibitem{csimcsek2016semantic}
U.~{\c{S}}im{\c{s}}ek, A.~Fensel, A.~Zafeiropoulos, E.~Fotopoulou, P.~Liapis,
  T.~Bouras, F.~T. Saenz, A.~F.~S. G{\'o}mez, A semantic approach towards
  implementing energy efficient lifestyles through behavioural change, in:
  Proceedings of the 12th International Conference on Semantic Systems, 2016,
  pp. 173--176.

\bibitem{Varlamis2020CCIS}
I.~Varlamis, C.~Sardianos, G.~Dimitrakopoulos, A.~Alsalemi, Y.~Himeur,
  F.~Bensaali, A.~Amira, Reshaping consumption habits by exploiting
  energy-related micro-moment recommendations: A case study, in: Communications
  in Computer and Information Science, Springer International Publishing, Cham,
  2020, pp. 1--22.

\bibitem{iweka2019energy}
O.~Iweka, S.~Liu, A.~Shukla, D.~Yan, Energy and behaviour at home: a review of
  intervention methods and practices, Energy Research \& Social Science 57
  (2019) 101238.

\bibitem{sardianos2020emergence}
C.~Sardianos, I.~Varlamis, C.~Chronis, G.~Dimitrakopoulos, A.~Alsalemi,
  Y.~Himeur, F.~Bensaali, A.~Amira, The emergence of explainability of
  intelligent systems: Delivering explainable and personalized recommendations
  for energy efficiency, International Journal of Intelligent Systems 36~(2)
  (2021) 656--680.

\bibitem{kitchenham2004procedures}
B.~Kitchenham, Procedures for performing systematic reviews, Keele, UK, Keele
  University 33~(2004) (2004) 1--26.

\bibitem{hauge2011user}
{\AA}.~L. Hauge, J.~Thomsen, T.~Berker, User evaluations of energy efficient
  buildings: Literature review and further research, Advances in Building
  Energy Research 5~(1) (2011) 109--127.

\bibitem{taherahmadi2020toward}
J.~Taherahmadi, Y.~Noorollahi, M.~Panahi, Toward comprehensive zero energy
  building definitions: a literature review and recommendations, International
  Journal of Sustainable Energy (2020) 1--29.

\bibitem{boodi2018intelligent}
A.~Boodi, K.~Beddiar, M.~Benamour, Y.~Amirat, M.~Benbouzid, Intelligent systems
  for building energy and occupant comfort optimization: A state of the art
  review and recommendations, Energies 11~(10) (2018) 2604.

\bibitem{de2014intelligent}
A.~De~Paola, M.~Ortolani, G.~Lo~Re, G.~Anastasi, S.~K. Das, Intelligent
  management systems for energy efficiency in buildings: A survey, ACM
  Computing Surveys (CSUR) 47~(1) (2014) 1--38.

\bibitem{khajenasiri2017review}
I.~Khajenasiri, A.~Estebsari, M.~Verhelst, G.~Gielen, A review on internet of
  things solutions for intelligent energy control in buildings for smart city
  applications, Energy Procedia 111 (2017) 770--779.

\bibitem{shareef2018review}
H.~Shareef, M.~S. Ahmed, A.~Mohamed, E.~Al~Hassan, Review on home energy
  management system considering demand responses, smart technologies, and
  intelligent controllers, Ieee Access 6 (2018) 24498--24509.

\bibitem{resnick1997recommender}
P.~Resnick, H.~R. Varian, Recommender systems, Communications of the ACM 40~(3)
  (1997) 56--58.

\bibitem{schafer1999recommender}
J.~B. Schafer, J.~Konstan, J.~Riedl, Recommender systems in e-commerce, in:
  Proceedings of the 1st ACM conference on Electronic commerce, 1999, pp.
  158--166.

\bibitem{martin2009recsys}
F.~J. Martin, Recsys' 09 industrial keynote: top 10 lessons learned developing
  deploying and operating real-world recommender systems, in: Proceedings of
  the third ACM conference on Recommender systems, 2009, pp. 1--2.

\bibitem{bao2015recommendations}
J.~Bao, Y.~Zheng, D.~Wilkie, M.~Mokbel, Recommendations in location-based
  social networks: a survey, GeoInformatica 19~(3) (2015) 525--565.

\bibitem{alsalemi2020achieving}
A.~Alsalemi, Y.~Himeur, F.~Bensaali, A.~Amira, C.~Sardianos, I.~Varlamis,
  G.~Dimitrakopoulos, Achieving domestic energy efficiency using micro-moments
  and intelligent recommendations, IEEE Access 8 (2020) 15047--15055.

\bibitem{sardianos2019want}
C.~Sardianos, I.~Varlamis, G.~Dimitrakopoulos, D.~Anagnostopoulos, A.~Alsalemi,
  F.~Bensaali, A.~Amira, " i want to... change": Micro-moment based
  recommendations can change users' energy habits., in: SMARTGREENS, 2019, pp.
  30--39.

\bibitem{eirinaki2018recommender}
M.~Eirinaki, J.~Gao, I.~Varlamis, K.~Tserpes, Recommender systems for
  large-scale social networks: A review of challenges and solutions, Future
  Generation Computer Systems 78 (2018) 413--418.

\bibitem{castells2015novelty}
P.~Castells, N.~J. Hurley, S.~Vargas, Novelty and diversity in recommender
  systems, in: Recommender systems handbook, Springer, 2015, pp. 881--918.

\bibitem{Pinto2018}
T.~Pinto, R.~Faia, M.~Navarro, G.~Santos, J.~Corchado~Rodríguez, Z.~Vale,
  Multi-agent-based cbr recommender system for intelligent energy management in
  buildings, IEEE Systems Journal PP (2018) 1--12.

\bibitem{kaur2019energy}
E.~Kaur, S.~Sharma, A.~Verma, M.~Singh, An energy management and recommender
  system for lighting control in internet-of-energy enabled buildings,
  IFAC-PapersOnLine 52~(4) (2019) 288--293.

\bibitem{cui2016short}
C.~Cui, T.~Wu, M.~Hu, J.~D. Weir, X.~Li, Short-term building energy model
  recommendation system: A meta-learning approach, Applied energy 172 (2016)
  251--263.

\bibitem{cuffaro2017resource}
G.~Cuffaro, F.~Paganelli, G.~Mylonas, A resource-based rule engine for energy
  savings recommendations in educational buildings, in: 2017 Global Internet of
  Things Summit (GIoTS), IEEE, 2017, pp. 1--6.

\bibitem{Wei2018}
P.~Wei, S.~Xia, X.~Jiang, Energy saving recommendations and user location
  modeling in commercial buildings, in: Proceedings of the 26th Conference on
  User Modeling, Adaptation and Personalization, UMAP ’18, Association for
  Computing Machinery, ACM, New York, NY, USA, 2018, p. 3–11.

\bibitem{wei2020deep}
P.~Wei, S.~Xia, R.~Chen, J.~Qian, C.~Li, X.~Jiang, A deep reinforcement
  learning based recommender system for occupant-driven energy optimization in
  commercial buildings, IEEE Internet of Things Journal.

\bibitem{KAR2019135}
P.~Kar, A.~Shareef, A.~Kumar, K.~T. Harn, B.~Kalluri, S.~K. Panda, Revicee: A
  recommendation based approach for personalized control, visual comfort \&
  energy efficiency in buildings, Building and Environment 152 (2019) 135 --
  144.

\bibitem{Schweizer7424470}
D.~{Schweizer}, M.~{Zehnder}, H.~{Wache}, H.~{Witschel}, D.~{Zanatta},
  M.~{Rodriguez}, Using consumer behavior data to reduce energy consumption in
  smart homes: Applying machine learning to save energy without lowering
  comfort of inhabitants, in: 2015 IEEE 14th International Conference on
  Machine Learning and Applications (ICMLA), 2015, pp. 1123--1129.

\bibitem{OSADCHIY2019535}
T.~Osadchiy, I.~Poliakov, P.~Olivier, M.~Rowland, E.~Foster, Recommender system
  based on pairwise association rules, Expert Systems with Applications 115
  (2019) 535 -- 542.

\bibitem{dahihande2020reducing}
J.~Dahihande, A.~Jaiswal, A.~A. Pagar, A.~Thakare, M.~Eirinaki, I.~Varlamis,
  Reducing energy waste in households through real-time recommendations, in:
  Fourteenth ACM Conference on Recommender Systems, 2020, pp. 545--550.

\bibitem{morawski2017fuzzy}
J.~Morawski, T.~Stepan, S.~Dick, J.~Miller, A fuzzy recommender system for
  public library catalogs, International Journal of Intelligent Systems 32~(10)
  (2017) 1062--1084.

\bibitem{castro2018group}
J.~Castro, M.~J. Barranco, R.~M. Rodr{\'\i}guez, L.~Mart{\'\i}nez, Group
  recommendations based on hesitant fuzzy sets, International Journal of
  Intelligent Systems 33~(10) (2018) 2058--2077.

\bibitem{Zhang8412100}
Y.~{Zhang}, K.~{Meng}, W.~{Kong}, Z.~Y. {Dong}, Collaborative filtering-based
  electricity plan recommender system, IEEE Transactions on Industrial
  Informatics 15~(3) (2019) 1393--1404.

\bibitem{ZHENG2020117775}
J.~Zheng, C.~S. Lai, H.~Yuan, Z.~Y. Dong, K.~Meng, L.~L. Lai, Electricity plan
  recommender system with electrical instruction-based recovery, Energy 203
  (2020) 117775.

\bibitem{Adomavicius2015}
G.~Adomavicius, A.~Tuzhilin, Context-aware recommender systems, in: Recommender
  systems handbook, Springer, 2011, pp. 217--253.

\bibitem{sardianos2020model}
C.~Sardianos, I.~Varlamis, C.~Chronis, G.~Dimitrakopoulos, Y.~Himeur,
  A.~Alsalemi, F.~Bensaali, A.~Amira, A model for predicting room occupancy
  based on motion sensor data, in: 2020 IEEE International Conference on
  Informatics, IoT, and Enabling Technologies (ICIoT), IEEE, 2020, pp.
  394--399.

\bibitem{RAZA201984}
S.~Raza, C.~Ding, Progress in context-aware recommender systems — an
  overview, Computer Science Review 31 (2019) 84 -- 97.

\bibitem{alsalemi2020micro}
A.~Alsalemi, Y.~Himeur, F.~Bensaali, A.~Amira, C.~Sardianos, C.~Chronis,
  I.~Varlamis, G.~Dimitrakopoulos, A micro-moment system for domestic energy
  efficiency analysis, IEEE Systems Journal.

\bibitem{Shigeyoshi2013}
H.~Shigeyoshi, K.~Tamano, R.~Saga, H.~Tsuji, S.~Inoue, T.~Ueno, Social
  experiment on advisory recommender system for energy-saving, in: S.~Yamamoto
  (Ed.), Human Interface and the Management of Information. Information and
  Interaction Design, Springer, Springer Berlin Heidelberg, Berlin, Heidelberg,
  2013, pp. 545--554.

\bibitem{radha2016lifestyle}
M.~Radha, M.~C. Willemsen, M.~Boerhof, W.~A. IJsselsteijn, Lifestyle
  recommendations for hypertension through rasch-based feasibility modeling,
  in: Proceedings of the 2016 Conference on User Modeling Adaptation and
  Personalization, 2016, pp. 239--247.

\bibitem{Starke2015}
A.~Starke, M.~Willemsen, C.~Snijders, Saving energy in 1-d: Tailoring
  energy-saving advice using a {R}asch-based energy recommender system, in:
  Proceedings of 2nd International Workshop on Decision Making and Recommender
  Systems, CEUR Workshop Proceedings, 2015, pp. 1--8.

\bibitem{Starke2017}
A.~Starke, M.~Willemsen, C.~Snijders, Effective user interface designs to
  increase energy-efficient behavior in a rasch-based energy recommender
  system, in: Proceedings of the Eleventh ACM Conference on Recommender
  Systems, RecSys ’17, ACM, Association for Computing Machinery, New York,
  NY, USA, 2017, p. 65–73.

\bibitem{chulyadyo2014personalized}
R.~Chulyadyo, P.~Leray, A personalized recommender system from probabilistic
  relational model and users’ preferences, Procedia Computer Science 35
  (2014) 1063--1072.

\bibitem{kumar2001recommendation}
R.~Kumar, P.~Raghavan, S.~Rajagopalan, A.~Tomkins, Recommendation systems: A
  probabilistic analysis, Journal of Computer and System Sciences 63~(1) (2001)
  42--61.

\bibitem{Li7093924}
L.~Kung, H.-F. Wang, A recommender system for the optimal combination of energy
  resources with cost-benefit analysis, in: 2015 International Conference on
  Industrial Engineering and Operations Management (IEOM), IEEE, 2015, pp.
  1--10.

\bibitem{OKU2011}
K.~Oku, F.~Hattori, Fusion-based recommender system for improving serendipity,
  CEUR Workshop Proceedings 816 (2011) 19--26.

\bibitem{xin2011effective}
X.~Xin, Effective fusion-based approaches for recommender systems, Chinese
  University of Hong Kong, 2011.

\bibitem{zhang2010fusion}
K.~Zhang, H.~Li, Fusion-based recommender system, in: 2010 13th International
  Conference on Information Fusion, IEEE, 2010, pp. 1--7.

\bibitem{himeur2020data}
Y.~Himeur, A.~Alsalemi, A.~Al-Kababji, F.~Bensaali, A.~Amira, Data fusion
  strategies for energy efficiency in buildings: Overview, challenges and novel
  orientations, Information Fusion 64 (2020) 99--120.

\bibitem{wroblewska2020multimodal}
A.~Wroblewska, J.~Dabrowski, M.~Pastuszak, A.~Michalowski, M.~Daniluk,
  B.~Rychalska, M.~Wieczorek, S.~Sysko-Romanczuk, Multi-modal embedding
  fusion-based recommender (2020).
\newblock \href {http://arxiv.org/abs/2005.06331} {\path{arXiv:2005.06331}}.

\bibitem{ji2020brs}
Z.~Ji, C.~Yang, H.~Wang, J.~E. Armend{\'a}riz-i{\~n}igo, M.~Arce-Urriza, Brs c
  s: a hybrid recommendation model fusing multi-source heterogeneous data,
  EURASIP Journal on Wireless Communications and Networking 2020~(1) (2020)
  1--17.

\bibitem{shambour2012trust}
Q.~Shambour, J.~Lu, A trust-semantic fusion-based recommendation approach for
  e-business applications, Decision Support Systems 54~(1) (2012) 768--780.

\bibitem{wang2019collaborative}
H.~Wang, Y.~Song, P.~Mi, J.~Duan, The collaborative filtering method based on
  social information fusion, Mathematical Problems in Engineering 2019.

\bibitem{pradhan2020multi}
T.~Pradhan, S.~Pal, A multi-level fusion based decision support system for
  academic collaborator recommendation, Knowledge-Based Systems (2020) 105784.

\bibitem{aiello2018decision}
G.~Aiello, I.~Giovino, M.~Vallone, P.~Catania, A.~Argento, A decision support
  system based on multisensor data fusion for sustainable greenhouse
  management, Journal of Cleaner Production 172 (2018) 4057--4065.

\bibitem{HimeurCOGN2020}
Y.~Himeur, A.~Alsalemi, F.~Bensaali, A.~Amira, A novel approach for detecting
  anomalous energy consumption based on micro-moments and deep neural networks,
  Cognitive Computation 12~(6) (2020) 1381--1401.

\bibitem{Zhang10.1145/3285029}
S.~Zhang, L.~Yao, A.~Sun, Y.~Tay, Deep learning based recommender system: A
  survey and new perspectives, ACM Computing Surveys (CSUR) 52~(1) (2019)
  1--38.

\bibitem{app10072441}
J.~Bobadilla, S.~Alonso, A.~Hernando, Deep learning architecture for
  collaborative filtering recommender systems, Applied Sciences 10~(7).

\bibitem{R2020113054}
K.~R, P.~Kumar, B.~Bhasker, Dnnrec: A novel deep learning based hybrid
  recommender system, Expert Systems with Applications 144 (2020) 113054.

\bibitem{app10144926}
R.~Lara-Cabrera, {\'A}.~Gonz{\'a}lez-Prieto, F.~Ortega, Deep matrix
  factorization approach for collaborative filtering recommender systems,
  Applied Sciences 10~(14) (2020) 4926.

\bibitem{shah2019review}
A.~S. Shah, H.~Nasir, M.~Fayaz, A.~Lajis, A.~Shah, A review on energy
  consumption optimization techniques in iot based smart building environments,
  Information 10~(3) (2019) 108.

\bibitem{tushar2019optimizing}
Q.~Tushar, M.~Bhuiyan, M.~Sandanayake, G.~Zhang, Optimizing the energy
  consumption in a residential building at different climate zones: Towards
  sustainable decision making, Journal of cleaner production 233 (2019)
  634--649.

\bibitem{ceballos2019simulation}
I.~Ceballos-Fuentealba, E.~{\'A}lvarez-Miranda, C.~Torres-Fuchslocher, M.~L.
  del Campo-Hitschfeld, J.~D{\'\i}az-Guerrero, A simulation and optimisation
  methodology for choosing energy efficiency measures in non-residential
  buildings, Applied Energy 256 (2019) 113953.

\bibitem{rocha2015improving}
P.~Rocha, A.~Siddiqui, M.~Stadler, Improving energy efficiency via smart
  building energy management systems: A comparison with policy measures, Energy
  and Buildings 88 (2015) 203--213.

\bibitem{anvari2017multi}
A.~Anvari-Moghaddam, A.~Rahimi-Kian, M.~S. Mirian, J.~M. Guerrero, A
  multi-agent based energy management solution for integrated buildings and
  microgrid system, Applied energy 203 (2017) 41--56.

\bibitem{lu2020economic}
S.~Lu, W.~Gu, K.~Meng, Z.~Y. Dong, Economic dispatch of integrated energy
  systems with robust thermal comfort management, IEEE Transactions on
  Sustainable Energy.

\bibitem{di2017two}
M.~Di~Piazza, G.~La~Tona, M.~Luna, A.~Di~Piazza, A two-stage energy management
  system for smart buildings reducing the impact of demand uncertainty, Energy
  and Buildings 139 (2017) 1--9.

\bibitem{salakij2016model}
S.~Salakij, N.~Yu, S.~Paolucci, P.~Antsaklis, Model-based predictive control
  for building energy management. i: Energy modeling and optimal control,
  Energy and Buildings 133 (2016) 345--358.

\bibitem{yu2017model}
N.~Yu, S.~Salakij, R.~Chavez, S.~Paolucci, M.~Sen, P.~Antsaklis, Model-based
  predictive control for building energy management: Part ii--experimental
  validations, Energy and Buildings 146 (2017) 19--26.

\bibitem{paul2019real}
S.~Paul, N.~P. Padhy, Real-time bilevel energy management of smart residential
  apartment building, IEEE Transactions on Industrial Informatics 16~(6) (2019)
  3708--3720.

\bibitem{lu2019thermal}
S.~Lu, W.~Gu, K.~Meng, S.~Yao, B.~Liu, Z.~Y. Dong, Thermal inertial aggregation
  model for integrated energy systems, IEEE Transactions on Power Systems
  35~(3) (2019) 2374--2387.

\bibitem{abbasi2019software}
A.~A. Abbasi, A.~Abbasi, S.~Shamshirband, A.~T. Chronopoulos, V.~Persico,
  A.~Pescap{\`e}, Software-defined cloud computing: A systematic review on
  latest trends and developments, IEEE Access 7 (2019) 93294--93314.

\bibitem{AlsalemiRTDPCC2020}
A.~{Elsalemi}, A.~{Al-kababji}, Y.~{Himeur}, F.~{Bensaali}, A.~Amira, Cloud
  energy micro-moment data classification: A platform study, in: IEEE
  SmartWorld, Ubiquitous Intelligence Computing, Advanced Trusted Computing,
  Scalable Computing Communications, Cloud Big Data Computing, Internet of
  People and Smart City Innovation (SmartWorld/SCALCOM/UIC/ATC/CBDCom/IOP/SCI),
  2020, pp. 1--6.

\bibitem{zhou2017security}
J.~Zhou, Z.~Cao, X.~Dong, A.~V. Vasilakos, Security and privacy for cloud-based
  iot: Challenges, IEEE Communications Magazine 55~(1) (2017) 26--33.

\bibitem{ren2019survey}
J.~Ren, D.~Zhang, S.~He, Y.~Zhang, T.~Li, A survey on end-edge-cloud
  orchestrated network computing paradigms: Transparent computing, mobile edge
  computing, fog computing, and cloudlet, ACM Computing Surveys (CSUR) 52~(6)
  (2019) 1--36.

\bibitem{Alsalaemi2020sca}
A.~{Elsalemi}, Y.~{Himeur}, F.~{Bensaali}, A.~Amira, Appliance-level monitoring
  with micro-moment smart plugs, in: The Fifth International Conference on
  Smart City Applications (SCA), 2020, pp. 1--5.

\bibitem{HIMEUR2020115872}
Y.~Himeur, A.~Alsalemi, F.~Bensaali, A.~Amira, Effective non-intrusive load
  monitoring of buildings based on a novel multi-descriptor fusion with
  dimensionality reduction, Applied Energy 279 (2020) 115872.

\bibitem{Himeur2020icpr}
Y.~{Himeur}, A.~{Elsalemi}, F.~{Bensaali}, A.~Amira, Appliance identification
  using a histogram post-processing of 2d local binary patterns for smart grid
  applications, in: Proc. 25th International Conference on Pattern Recognition
  (ICPR), 2020, pp. 1--8.

\bibitem{devarajan2019fog}
M.~Devarajan, V.~Subramaniyaswamy, V.~Vijayakumar, L.~Ravi, Fog-assisted
  personalized healthcare-support system for remote patients with diabetes,
  Journal of Ambient Intelligence and Humanized Computing 10~(10) (2019)
  3747--3760.

\bibitem{hernandez2020fog}
E.~Hern{\'a}ndez-Nieves, G.~Hern{\'a}ndez, A.-B. Gil-Gonz{\'a}lez,
  S.~Rodr{\'\i}guez-Gonz{\'a}lez, J.~M. Corchado, Fog computing architecture
  for personalized recommendation of banking products, Expert Systems with
  Applications 140 (2020) 112900.

\bibitem{wang2019fog}
X.~Wang, B.~Gu, Y.~Ren, W.~Ye, S.~Yu, Y.~Xiang, L.~Gao, A fog-based recommender
  system, IEEE Internet of Things Journal 7~(2) (2019) 1048--1060.

\bibitem{linthicum_edge_nodate}
D.~Linthicum,
  \href{https://techbeacon.com/enterprise-it/edge-computing-hybrid-cloud-3-approaches}{Edge
  computing in hybrid cloud: 3 approaches} (2019).
\url{https://techbeacon.com/enterprise-it/edge-computing-hybrid-cloud-3-approaches}

\bibitem{huang2019deepar}
Y.~Huang, F.~Wang, F.~Wang, J.~Liu, Deepar: A hybrid device-edge-cloud
  execution framework for mobile deep learning applications, in: IEEE INFOCOM
  2019-IEEE Conference on Computer Communications Workshops (INFOCOM WKSHPS),
  IEEE, 2019, pp. 892--897.

\bibitem{gunawardana2015evaluating}
A.~Gunawardana, G.~Shani, {Evaluating Recommender Systems}, in: Recommender
  systems handbook, Springer, 2015, pp. 265--308.

\bibitem{wu2012evaluating}
W.~Wu, L.~He, J.~Yang, Evaluating recommender systems, in: Seventh
  International Conference on Digital Information Management (ICDIM 2012),
  IEEE, 2012, pp. 56--61.

\bibitem{chai2014root}
T.~Chai, R.~R. Draxler, Root mean square error (rmse) or mean absolute error
  (mae)?--arguments against avoiding rmse in the literature, Geoscientific
  model development 7~(3) (2014) 1247--1250.

\bibitem{herlocker2004evaluating}
J.~L. Herlocker, J.~A. Konstan, L.~G. Terveen, J.~T. Riedl, Evaluating
  collaborative filtering recommender systems, ACM Transactions on Information
  Systems (TOIS) 22~(1) (2004) 5--53.

\bibitem{herlocker2002empirical}
J.~Herlocker, J.~A. Konstan, J.~Riedl, An empirical analysis of design choices
  in neighborhood-based collaborative filtering algorithms, Information
  retrieval 5~(4) (2002) 287--310.

\bibitem{mcnee2006making}
S.~M. McNee, J.~Riedl, J.~A. Konstan, {Making Recommendations Better: An
  Analytic Model for Human-Recommender Interaction}, in: CHI'06 extended
  abstracts on Human factors in computing systems, 2006, pp. 1103--1108.

\bibitem{karjalainen2011consumer}
S.~Karjalainen, Consumer preferences for feedback on household electricity
  consumption, Energy and buildings 43~(2-3) (2011) 458--467.

\bibitem{karlin2015effects}
B.~Karlin, J.~F. Zinger, R.~Ford, The effects of feedback on energy
  conservation: A meta-analysis., Psychological bulletin 141~(6) (2015) 1205.

\bibitem{garbi2019beneffice}
A.~Garbi, A.~Malamou, N.~Michas, Z.~Pontikas, N.~Doulamis, E.~Protopapadakis,
  T.~N. Mikkelsen, K.~Kanellakis, J.-L. Baradat, Beneffice: Behaviour change,
  consumption monitoring and analytics with complementary currency rewards,
  Proceedings 20~(1) (2019) 12.

\bibitem{petkov2011engaging}
P.~Petkov, F.~K{\"o}bler, M.~Foth, R.~Medland, H.~Krcmar, Engaging energy
  saving through motivation-specific social comparison, in: CHI'11 Extended
  Abstracts on Human Factors in Computing Systems, Association for Computing
  Machinery, 2011, pp. 1945--1950.

\bibitem{jain2013can}
R.~K. Jain, R.~Gulbinas, J.~E. Taylor, P.~J. Culligan, Can social influence
  drive energy savings? detecting the impact of social influence on the energy
  consumption behavior of networked users exposed to normative eco-feedback,
  Energy and Buildings 66 (2013) 119--127.

\bibitem{du2016modelling}
F.~Du, J.~Zhang, H.~Li, J.~Yan, S.~Galloway, K.~L. Lo, Modelling the impact of
  social network on energy savings, Applied Energy 178 (2016) 56--65.

\bibitem{wemyss2019does}
D.~Wemyss, F.~Cellina, E.~Lobsiger-K{\"a}gi, V.~De~Luca, R.~Castri, Does it
  last? long-term impacts of an app-based behavior change intervention on
  household electricity savings in switzerland, Energy Research \& Social
  Science 47 (2019) 16--27.

\bibitem{morley2018digitalisation}
J.~Morley, K.~Widdicks, M.~Hazas, Digitalisation, energy and data demand: The
  impact of internet traffic on overall and peak electricity consumption,
  Energy Research \& Social Science 38 (2018) 128--137.

\bibitem{Garcia2017}
O.~Garcia, J.~Prieto, R.~Alonso, J.~Corchado~Rodríguez, A framework to improve
  energy efficient behaviour at home through activity and context monitoring,
  Sensors 17 (2017) 1749.

\bibitem{Luo2017}
F.~Luo, G.~Ranzi, W.~Kong, Z.~Dong, S.~Wang, J.~Zhao, Non-intrusive energy
  saving appliance recommender system for smart grid residential users, IET
  Generation, Transmission and Distribution 11.

\bibitem{SARDIANOS2020394}
C.~Sardianos, I.~Varlamis, G.~Dimitrakopoulos, D.~Anagnostopoulos, A.~Alsalemi,
  F.~Bensaali, Y.~Himeur, A.~Amira, Rehab-c: Recommendations for energy habits
  change, Future Generation Computer Systems 112 (2020) 394 -- 407.

\bibitem{Bravo2019}
D.~Jiménez-Bravo, J.~Pérez-Marcos, D.~Hernández de~la Iglesia,
  G.~Villarrubia, J.~Paz, Multi-agent recommendation system for electrical
  energy optimization and cost saving in smart homes, Energies 12 (2019) 1317.

\bibitem{ari2019enabling}
A.~A.~A. Ari, O.~K. Ngangmo, C.~Titouna, O.~Thiare, A.~Mohamadou, A.~M.
  Gueroui, et~al., Enabling privacy and security in cloud of things:
  Architecture, applications, security \& privacy challenges, Applied Computing
  and Informatics.

\bibitem{schaefer2020management}
J.~L. Schaefer, J.~C.~M. Siluk, P.~S.~d. Carvalho, J.~Renes~Pinheiro, P.~S.
  Schneider, Management challenges and opportunities for energy cloud
  development and diffusion, Energies 13~(16) (2020) 4048.

\bibitem{mahjabin2017survey}
T.~Mahjabin, Y.~Xiao, G.~Sun, W.~Jiang, A survey of distributed
  denial-of-service attack, prevention, and mitigation techniques,
  International Journal of Distributed Sensor Networks 13~(12) (2017)
  1550147717741463.

\bibitem{himeur2020marketability}
Y.~Himeur, A.~Alsalemi, F.~Bensaali, A.~Amira, I.~Varlamis, G.~Bravos,
  C.~Sardianos, Dimitrakopoulos, Techno-economic analysis of building energy
  efficiency systems based on behavioral change: A case study of a novel
  micro-moments based solution, Applied Energy (2021) 1--25.

\bibitem{natarajan2020resolving}
S.~Natarajan, S.~Vairavasundaram, S.~Natarajan, A.~H. Gandomi, Resolving data
  sparsity and cold start problem in collaborative filtering recommender system
  using linked open data, Expert Systems with Applications 149 (2020) 113248.

\bibitem{jain2020efficient}
A.~F. Jain, S.~K. Vishwakarma, P.~Jain, An efficient collaborative recommender
  system for removing sparsity problem, in: ICT Analysis and Applications,
  Springer, 2020, pp. 131--141.

\bibitem{zhang2020alleviating}
F.~Zhang, S.~Qi, Q.~Liu, M.~Mao, A.~Zeng, Alleviating the data sparsity problem
  of recommender systems by clustering nodes in bipartite networks, Expert
  Systems with Applications (2020) 113346.

\bibitem{son2016dealing}
L.~H. Son, Dealing with the new user cold-start problem in recommender systems:
  A comparative review, Information Systems 58 (2016) 87--104.

\bibitem{lika2014facing}
B.~Lika, K.~Kolomvatsos, S.~Hadjiefthymiades, Facing the cold start problem in
  recommender systems, Expert Systems with Applications 41~(4) (2014)
  2065--2073.

\bibitem{liu2014promoting}
J.-H. Liu, T.~Zhou, Z.-K. Zhang, Z.~Yang, C.~Liu, W.-M. Li, Promoting
  cold-start items in recommender systems, PloS one 9~(12) (2014) e113457.

\bibitem{verbert2011dataset}
K.~Verbert, H.~Drachsler, N.~Manouselis, M.~Wolpers, R.~Vuorikari, E.~Duval,
  Dataset-driven research for improving recommender systems for learning, in:
  Proceedings of the 1st International Conference on Learning Analytics and
  Knowledge, 2011, pp. 44--53.

\bibitem{drachsler2010issues}
H.~Drachsler, T.~Bogers, R.~Vuorikari, K.~Verbert, E.~Duval, N.~Manouselis,
  G.~Beham, S.~Lindstaedt, H.~Stern, M.~Friedrich, et~al., Issues and
  considerations regarding sharable data sets for recommender systems in
  technology enhanced learning, Procedia Computer Science 1~(2) (2010)
  2849--2858.

\bibitem{ccano2015characterization}
E.~{\c{C}}ano, M.~Morisio, Characterization of public datasets for recommender
  systems, in: 2015 IEEE 1st International Forum on Research and Technologies
  for Society and Industry Leveraging a better tomorrow (RTSI), IEEE, 2015, pp.
  249--257.

\bibitem{isinkaye2015recommendation}
F.~Isinkaye, Y.~Folajimi, B.~A. Ojokoh, Recommendation systems: Principles,
  methods and evaluation, Egyptian informatics journal 16~(3) (2015) 261--273.

\bibitem{dacrema2019troubling}
M.~F. Dacrema, S.~Boglio, P.~Cremonesi, D.~Jannach, A troubling analysis of
  reproducibility and progress in recommender systems research, arXiv preprint
  arXiv:1911.07698.

\bibitem{ekstrand2011rethinking}
M.~D. Ekstrand, M.~Ludwig, J.~A. Konstan, J.~T. Riedl, Rethinking the
  recommender research ecosystem: reproducibility, openness, and lenskit, in:
  Proceedings of the fifth ACM conference on Recommender systems, 2011, pp.
  133--140.

\bibitem{HimeurENB2020}
Y.~Himeur, A.~Alsalemi, F.~Bensaali, A.~Amira, Building power consumption
  datasets: Survey, taxonomy and future directions, Energy and Buildings 227
  (2020) 110404.

\bibitem{beel2016towards}
J.~Beel, C.~Breitinger, S.~Langer, A.~Lommatzsch, B.~Gipp, Towards
  reproducibility in recommender-systems research, User modeling and
  user-adapted interaction 26~(1) (2016) 69--101.

\bibitem{ie2019recsim}
E.~Ie, C.~wei Hsu, M.~Mladenov, V.~Jain, S.~Narvekar, J.~Wang, R.~Wu,
  C.~Boutilier, Recsim: A configurable simulation platform for recommender
  systems (2019).
\newblock \href {http://arxiv.org/abs/1909.04847} {\path{arXiv:1909.04847}}.

\bibitem{trianni2013drivers}
A.~Trianni, E.~Cagno, F.~Marchesani, G.~Spallina, Drivers for industrial energy
  efficiency: an innovative approach, in: ICAE--International Conference on
  Applied Energy, 2013, pp. 1--9.

\bibitem{climateenergy2030}
O.~Fitch-Roy, J.~Fairbrass, Negotiating the EU’s 2030 climate and energy
  framework: Agendas, ideas and European interest groups, Springer, 2018.

\bibitem{kaabi2012conservation}
F.~Kaabi, Conservation plan for tarsheed, Qatar General Electricity and Water
  Corporation, Conservation \& Energy Efficiency Department, Doha, Qatar.

\bibitem{tarhseed_qatar_2020}
{Qatar} {General} {Electricity} \& water {Corporation}, Available online:
  \url{https://www.km.qa/Tarsheed/Pages/TarsheedIntro.aspx}, note = {Accessed:
  2020-11-05},.

\bibitem{gunning2017explainable}
D.~Gunning, Explainable artificial intelligence (xai), Defense Advanced
  Research Projects Agency (DARPA), nd Web 2~(2).

\bibitem{arrieta2020explainable}
A.~B. Arrieta, N.~D{\'\i}az-Rodr{\'\i}guez, J.~Del~Ser, A.~Bennetot, S.~Tabik,
  A.~Barbado, S.~Garc{\'\i}a, S.~Gil-L{\'o}pez, D.~Molina, R.~Benjamins,
  et~al., Explainable artificial intelligence (xai): Concepts, taxonomies,
  opportunities and challenges toward responsible ai, Information Fusion 58
  (2020) 82--115.

\bibitem{WU2021165}
Y.~Wu, Z.~Zhang, G.~Kou, H.~Zhang, X.~Chao, C.-C. Li, Y.~Dong, F.~Herrera,
  Distributed linguistic representations in decision making: Taxonomy, key
  elements and applications, and challenges in data science and explainable
  artificial intelligence, Information Fusion 65 (2021) 165 -- 178.

\bibitem{zhang2020explainable}
Y.~Zhang, X.~Chen, et~al., Explainable recommendation: A survey and new
  perspectives, Foundations and Trends{\textregistered} in Information
  Retrieval 14~(1) (2020) 1--101.

\bibitem{SAGI2020124}
O.~Sagi, L.~Rokach, Explainable decision forest: Transforming a decision forest
  into an interpretable tree, Information Fusion 61 (2020) 124 -- 138.

\bibitem{fernandez2020random}
R.~R. Fern{\'a}ndez, I.~M. de~Diego, V.~Ace{\~n}a, A.~Fern{\'a}ndez-Isabel,
  J.~M. Moguerza, Random forest explainability using counterfactual sets,
  Information Fusion 63 (2020) 196--207.

\bibitem{zhang2018explainable}
Y.~Zhang, X.~Chen, Explainable recommendation: A survey and new perspectives,
  arXiv preprint arXiv:1804.11192.

\bibitem{gao2019explainable}
J.~Gao, X.~Wang, Y.~Wang, X.~Xie, Explainable recommendation through attentive
  multi-view learning, in: Proceedings of the 33rd Conference on Artificial
  Intelligence, Vol.~33, AAAI, 2019, pp. 3622--3629.

\bibitem{balog2019transparent}
K.~Balog, F.~Radlinski, S.~Arakelyan, Transparent, scrutable and explainable
  user models for personalized recommendation, in: Proceedings of the 42nd
  International ACM SIGIR Conference on Research and Development in Information
  Retrieval, ACM, 2019, pp. 265--274.

\bibitem{catherine2017explainable}
R.~Catherine, K.~Mazaitis, M.~Eskenazi, W.~Cohen, Explainable entity-based
  recommendations with knowledge graphs, arXiv preprint arXiv:1707.05254.

\bibitem{zhang2014explicit}
Y.~Zhang, G.~Lai, M.~Zhang, Y.~Zhang, Y.~Liu, S.~Ma, Explicit factor models for
  explainable recommendation based on phrase-level sentiment analysis, in:
  Proceedings of the 37th International ACM SIGIR conference on Research \&
  Development in Information Retrieval, ACM, 2014, pp. 83--92.

\bibitem{he2015trirank}
X.~He, T.~Chen, M.-Y. Kan, X.~Chen, Trirank: Review-aware explainable
  recommendation by modeling aspects, in: Proceedings of the 24th ACM
  International on Conference on Information and Knowledge Management, ACM,
  2015, pp. 1661--1670.

\bibitem{grimaldo2019user}
A.~I. Grimaldo, J.~Novak, User-centered visual analytics approach for
  interactive and explainable energy demand analysis in prosumer scenarios, in:
  International Conference on Computer Vision Systems, Springer, 2019, pp.
  700--710.

\bibitem{sardianos2020data}
C.~Sardianos, I.~Varlamis, C.~Chronis, G.~Dimitrakopoulos, Y.~Himeur,
  A.~Alsalemi, F.~Bensaali, A.~Amira, Data analytics, automations, and
  micro-moment based recommendations for energy efficiency, in: 2020 IEEE Sixth
  International Conference on Big Data Computing Service and Applications
  (BigDataService), IEEE, 2020, pp. 96--103.

\bibitem{shi2015towards}
F.~Shi, J.-L. Marini, E.~Audry, Towards a psycho-cognitive recommender system,
  in: Proceedings of the International Workshop on Emotion Representations and
  Modelling for Companion Technologies, 2015, pp. 25--31.

\bibitem{zhao2014context}
L.~Zhao, J.~Huang, N.~Zhong, A context-aware recommender system with a
  cognition inspired model, in: International Conference on Rough Sets and
  Knowledge Technology, Springer, 2014, pp. 613--622.

\bibitem{shafto2016human}
P.~Shafto, O.~Nasraoui, Human-recommender systems: From benchmark data to
  benchmark cognitive models, in: Proceedings of the 10th ACM Conference on
  Recommender Systems, 2016, pp. 127--130.

\bibitem{kopeinik2016improving}
S.~Kopeinik, D.~Kowald, I.~Hasani-Mavriqi, E.~Lex, Improving collaborative
  filtering using a cognitive model of human category learning, The Journal of
  Web Science 2.

\bibitem{hamlabadi2017framework}
K.~G. HamlAbadi, A.~M. Saghiri, M.~Vahdati, M.~D. TakhtFooladi, M.~R. Meybodi,
  A framework for cognitive recommender systems in the internet of things
  (iot), in: 2017 IEEE 4th International Conference on Knowledge-Based
  Engineering and Innovation (KBEI), IEEE, 2017, pp. 0971--0976.

\bibitem{kopeinik2017applying}
S.~Kopeinik, Applying cognitive learner models for recommender systems in
  sparse data learning environments, SIGIR Forum 51~(3) (2018) 165.

\bibitem{aguilar2017general}
J.~Aguilar, P.~Valdiviezo-D{\'\i}az, G.~Riofrio, A general framework for
  intelligent recommender systems, Applied computing and informatics 13~(2)
  (2017) 147--160.

\bibitem{beheshti2020towards}
A.~Beheshti, S.~Yakhchi, S.~Mousaeirad, S.~M. Ghafari, S.~R. Goluguri, M.~A.
  Edrisi, Towards cognitive recommender systems, Algorithms 13~(8) (2020) 176.

\bibitem{bothos2016recommender}
E.~Bothos, D.~Apostolou, G.~Mentzas, A recommender for persuasive messages in
  route planning applications, in: 2016 7th International Conference on
  Information, Intelligence, Systems \& Applications (IISA), IEEE, 2016, pp.
  1--5.

\bibitem{sanchez2020persuasion}
R.~S{\'a}nchez-Corcuera, D.~Casado-Mansilla, C.~E. Borges, D.~L{\'o}pez-de
  Ipi{\~n}a, Persuasion-based recommender system ensambling matrix
  factorisation and active learning models, Personal and Ubiquitous Computing
  (2020) 1--11.

\bibitem{puglisi2015content}
S.~Puglisi, J.~Parra-Arnau, J.~Forn{\'e}, D.~Rebollo-Monedero, On content-based
  recommendation and user privacy in social-tagging systems, Computer Standards
  \& Interfaces 41 (2015) 17--27.

\bibitem{tang2017privacy}
Q.~Tang, H.~Wang, Privacy-preserving hybrid recommender system, in: Proceedings
  of the Fifth ACM International Workshop on Security in Cloud Computing, 2017,
  pp. 59--66.

\bibitem{shin2018privacy}
H.~Shin, S.~Kim, J.~Shin, X.~Xiao, Privacy enhanced matrix factorization for
  recommendation with local differential privacy, IEEE Transactions on
  Knowledge and Data Engineering 30~(9) (2018) 1770--1782.

\bibitem{li2017efficient}
D.~Li, Q.~Lv, L.~Shang, N.~Gu, Efficient privacy-preserving content
  recommendation for online social communities, Neurocomputing 219 (2017)
  440--454.

\bibitem{tseng2016privacy}
C.-M. Tseng, C.-K. Chau, On the privacy of crowd-sourced data collection for
  distance-to-empty prediction and eco-routing, in: Proceedings of the Workshop
  on Electric Vehicle Systems, Data, and Applications, 2016, pp. 1--6.

\bibitem{wang2016trust}
X.~Wang, J.~Zhang, Y.~Wang, Trust-aware privacy-preserving recommender system,
  in: Proceedings of the 9th EAI International Conference on Mobile Multimedia
  Communications, 2016, pp. 107--115.

\bibitem{mclaughlin2011protecting}
S.~McLaughlin, P.~McDaniel, W.~Aiello, Protecting consumer privacy from
  electric load monitoring, in: Proceedings of the 18th ACM conference on
  Computer and communications security, 2011, pp. 87--98.

\bibitem{himeur2020smart}
Y.~Himeur, A.~Alsalemi, F.~Bensaali, A.~Amira, Smart non-intrusive appliance
  identification using a novel local power histogramming descriptor with an
  improved k-nearest neighbors classifier, Sustainable Cities and Society 67
  (2021) 102764.

\bibitem{englert2015enhancing}
F.~Englert, M.~Rettberg-P{\"a}plow, S.~K{\"o}ssler, A.~Alhamoud, T.~A.~B.
  Nguyen, D.~B{\"o}hnstedt, R.~Steinmetz, Enhancing user privacy by data driven
  selection mechanisms for finding transmission-relevant data samples in energy
  recommender systems, in: 2015 International Conference and Workshops on
  Networked Systems (NetSys), IEEE, 2015, pp. 1--6.

\bibitem{badsha2017privacy}
S.~Badsha, X.~Yi, I.~Khalil, E.~Bertino, Privacy preserving user-based
  recommender system, in: 2017 IEEE 37th international conference on
  Distributed Computing Systems (ICDCS), IEEE, 2017, pp. 1074--1083.

\bibitem{jiang2019towards}
J.-Y. Jiang, C.-T. Li, S.-D. Lin, Towards a more reliable privacy-preserving
  recommender system, Information Sciences 482 (2019) 248--265.

\bibitem{badsha2016practical}
S.~Badsha, X.~Yi, I.~Khalil, A practical privacy-preserving recommender system,
  Data Science and Engineering 1~(3) (2016) 161--177.

\bibitem{xu2018privacy}
K.~Xu, W.~Zhang, Z.~Yan, A privacy-preserving mobile application recommender
  system based on trust evaluation, Journal of computational science 26 (2018)
  87--107.

\bibitem{begum2019towards}
N.~Begum, M.~Z.~A. Bhuiyan, Towards privacy-preserving recommender system with
  blockchains, in: Dependability in Sensor, Cloud, and Big Data Systems and
  Applications: 5th International Conference, DependSys 2019, Guangzhou, China,
  November 12--15, 2019, Proceedings, Vol. 1123, Springer Nature, 2019, p. 106.

\bibitem{pu2020efficient}
Y.~Pu, T.~Xiang, C.~Hu, A.~Alrawais, H.~Yan, An efficient blockchain-based
  privacy preserving scheme for vehicular social networks, Information Sciences
  540 (2020) 308--324.

\bibitem{bosri2020integrating}
R.~Bosri, M.~S. Rahman, M.~Z.~A. Bhuiyan, A.~Al~Omar, Integrating blockchain
  with artificial intelligence for privacy-preserving in recommender systems,
  IEEE Transactions on Network Science and Engineering.

\bibitem{casino2019efficient}
F.~Casino, C.~Patsakis, An efficient blockchain-based privacy-preserving
  collaborative filtering architecture, IEEE Transactions on Engineering
  Management.

\bibitem{gantner2010factorization}
Z.~Gantner, S.~Rendle, L.~Schmidt-Thieme, Factorization models for
  context-/time-aware movie recommendations, in: Proceedings of the Workshop on
  Context-Aware Movie Recommendation, 2010, pp. 14--19.

\bibitem{yuan2013time}
Q.~Yuan, G.~Cong, Z.~Ma, A.~Sun, N.~M. Thalmann, Time-aware point-of-interest
  recommendation, in: Proceedings of the 36th international ACM SIGIR
  conference on Research and development in information retrieval, 2013, pp.
  363--372.

\bibitem{zhang2015ticrec}
J.-D. Zhang, C.-Y. Chow, Ticrec: A probabilistic framework to utilize temporal
  influence correlations for time-aware location recommendations, IEEE
  Transactions on Services Computing 9~(4) (2015) 633--646.

\bibitem{stefanidis2013framework}
K.~Stefanidis, E.~Ntoutsi, M.~Petropoulos, K.~N{\o}rv{\aa}g, H.-P. Kriegel, A
  framework for modeling, computing and presenting time-aware recommendations,
  in: Transactions on Large-Scale Data-and Knowledge-Centered Systems X,
  Springer, 2013, pp. 146--172.

\bibitem{campos2014time}
P.~G. Campos, F.~D{\'\i}ez, I.~Cantador, Time-aware recommender systems: a
  comprehensive survey and analysis of existing evaluation protocols, User
  Modeling and User-Adapted Interaction 24~(1-2) (2014) 67--119.

\bibitem{linda2020effective}
S.~Linda, S.~Minz, K.~Bharadwaj, Effective context-aware recommendations based
  on context weighting using genetic algorithm and alleviating data sparsity,
  Applied Artificial Intelligence 34~(10) (2020) 730--753.

\bibitem{wang2018personalized}
K.~Wang, Y.~Jin, H.~Wang, H.~Peng, X.~Wang, Personalized time-aware tag
  recommendation, in: The Thirty-Second AAAI Conference on Artificial
  Intelligence (AAAI-18), 2018, pp. 459--466.

\bibitem{qian2019ears}
Y.~Qian, Y.~Zhang, X.~Ma, H.~Yu, L.~Peng, Ears: Emotion-aware recommender
  system based on hybrid information fusion, Information Fusion 46 (2019)
  141--146.

\bibitem{nilashi2020intelligent}
M.~Nilashi, S.~Asadi, R.~A. Abumalloh, S.~Samad, O.~Ibrahim, Intelligent
  recommender systems in the covid-19 outbreak: The case of wearable healthcare
  devices, Journal of Soft Computing and Decision Support Systems 7~(4) (2020)
  8--12.

\bibitem{lamche2015context}
B.~Lamche, Y.~R{\"o}dl, C.~Hauptmann, W.~W{\"o}rndl, Context-aware
  recommendations for mobile shopping., in: LocalRec@ RecSys, 2015, pp. 21--27.

\bibitem{unger2020context}
M.~Unger, A.~Tuzhilin, A.~Livne, Context-aware recommendations based on deep
  learning frameworks, ACM Transactions on Management Information Systems
  (TMIS) 11~(2) (2020) 1--15.

\bibitem{sanchez2018time}
P.~S{\'a}nchez, A.~Bellog{\'\i}n, Time-aware novelty metrics for recommender
  systems, in: European Conference on Information Retrieval, Springer, 2018,
  pp. 357--370.

\bibitem{kunaver2017diversity}
M.~Kunaver, T.~Po{\v{z}}rl, Diversity in recommender systems--a survey,
  Knowledge-Based Systems 123 (2017) 154--162.

\bibitem{batmaz2019review}
Z.~Batmaz, A.~Yurekli, A.~Bilge, C.~Kaleli, A review on deep learning for
  recommender systems: challenges and remedies, Artificial Intelligence Review
  52~(1) (2019) 1--37.

\bibitem{ying2018graph}
R.~Ying, R.~He, K.~Chen, P.~Eksombatchai, W.~L. Hamilton, J.~Leskovec, Graph
  convolutional neural networks for web-scale recommender systems, in:
  Proceedings of the 24th ACM SIGKDD International Conference on Knowledge
  Discovery \& Data Mining, 2018, pp. 974--983.

\bibitem{liu2016large}
C.-L. Liu, X.-W. Wu, Large-scale recommender system with compact latent factor
  model, Expert Systems with Applications 64 (2016) 467--475.

\bibitem{barraza2020introduction}
A.~Barraza-Urbina, D.~Glowacka, Introduction to bandits in recommender systems,
  in: Fourteenth ACM Conference on Recommender Systems, 2020, pp. 748--750.

\bibitem{zhou2017large}
Q.~Zhou, X.~Zhang, J.~Xu, B.~Liang, Large-scale bandit approaches for
  recommender systems, in: International Conference on Neural Information
  Processing, Springer, 2017, pp. 811--821.

\bibitem{canamares2019multi}
R.~Ca{\~n}amares, M.~Redondo, P.~Castells, Multi-armed recommender system
  bandit ensembles, in: Proceedings of the 13th ACM Conference on Recommender
  Systems, 2019, pp. 432--436.

\bibitem{sanz2019simple}
J.~Sanz-Cruzado, P.~Castells, E.~L{\'o}pez, A simple multi-armed
  nearest-neighbor bandit for interactive recommendation, in: Proceedings of
  the 13th ACM Conference on Recommender Systems, 2019, pp. 358--362.

\bibitem{sardianos2017scaling}
C.~Sardianos, I.~Varlamis, M.~Eirinaki, Scaling collaborative filtering to
  large--scale bipartite rating graphs using lenskit and spark, in: 2017 IEEE
  Third International Conference on Big Data Computing Service and Applications
  (BigDataService), IEEE, 2017, pp. 70--79.

\bibitem{bathla2020scalable}
G.~Bathla, H.~Aggarwal, R.~Rani, Scalable recommendation using large scale
  graph partitioning with pregel and giraph, International Journal of Cognitive
  Informatics and Natural Intelligence (IJCINI) 14~(4) (2020) 42--61.

\bibitem{li2019efficient}
H.~Li, K.~Li, J.~An, W.~Zheng, K.~Li, An efficient manifold regularized sparse
  non-negative matrix factorization model for large-scale recommender systems
  on gpus, Information Sciences 496 (2019) 464--484.

\bibitem{li2017msgd}
H.~Li, K.~Li, J.~An, K.~Li, Msgd: A novel matrix factorization approach for
  large-scale collaborative filtering recommender systems on gpus, IEEE
  Transactions on Parallel and Distributed Systems 29~(7) (2017) 1530--1544.

\bibitem{sardianos2018survey}
C.~Sardianos, N.~Tsirakis, I.~Varlamis, A survey on the scalability of
  recommender systems for social networks, in: Social Networks Science: Design,
  Implementation, Security, and Challenges, Springer, 2018, pp. 89--110.

\bibitem{zhang2018big}
Y.~Zhang, W.~Kong, Z.~Y. Dong, K.~Meng, J.~Qiu, Big data-driven electricity
  plan recommender system, in: 2018 IEEE Power \& Energy Society General
  Meeting (PESGM), IEEE, 2018, pp. 1--5.

\bibitem{kar2019revicee}
P.~Kar, A.~Shareef, A.~Kumar, K.~T. Harn, B.~Kalluri, S.~K. Panda, Revicee: A
  recommendation based approach for personalized control, visual comfort \&
  energy efficiency in buildings, Building and Environment 152 (2019) 135--144.

\bibitem{sardianos2020smart}
C.~Sardianos, C.~Chronis, I.~Varlamis, G.~Dimitrakopoulos, Y.~Himeur,
  A.~Alsalemi, F.~Bensaali, A.~Amira, Smart fusion of sensor data and human
  feedback for personalised energy-saving recommendations, International
  Journal of Intelligent Systems (2021) 1--20.

\bibitem{Fensel2013}
A.~Fensel, S.~Tomic, V.~Kumar, M.~Stefanovic, S.~V. Aleshin, D.~O. Novikov,
  {SESAME-S: Semantic Smart Home System for Energy Efficiency},
  Informatik-Spektrum 36~(1) (2013) 46--57.

\bibitem{RodriguezFernandez2016}
M.~{Rodr{\'{i}}guez Fern{\'{a}}ndez}, A.~{Cort{\'{e}}s Garc{\'{i}}a},
  I.~{Gonz{\'{a}}lez Alonso}, E.~{Zalama Casanova}, {Using the Big Data
  generated by the Smart Home to improve energy efficiency management}, Energy
  Effic. 9~(1) (2016) 249--260.

\bibitem{Fraternali2017}
P.~Fraternali, S.~Herrera, J.~Novak, M.~Melenhorst, D.~Tzovaras, S.~Krinidis,
  A.~E. Rizzoli, C.~Rottondi, F.~Cellina, {EnCOMPASS - An Integrative Approach
  to Behavioural Change for Energy Saving}, in: GIoTS 2017 - Glob. Internet
  Things Summit, Proc., 2017, pp. 1--6.

\bibitem{Fraternali2018}
P.~Fraternali, F.~Cellina, S.~Herrera, S.~Krinidis, C.~Pasini, A.~E. Rizzoli,
  C.~Rottondi, D.~Tzovaras, {A Socio-Technical System Based on Gamification
  Towards Energy Savings}, in: 2018 IEEE Int. Conf. Pervasive Comput. Commun.
  Work. (PerCom Work., IEEE, 2018, pp. 59--64.

\bibitem{Albertarelli2017}
S.~Albertarelli, P.~Fraternali, J.~Novak, A.~E. Rizzoli, C.~Rottondi, {DROP and
  FUNERGY: Two Gamified Learning Projects for Water and Energy Conservation},
  in: Proc. 11th Eur. Conf. Games Based Learn. ECGBL 2017, 2017, pp. 935--938.

\bibitem{Fotopoulou2017}
E.~Fotopoulou, A.~Zafeiropoulos, F.~Terroso, A.~Gonzalez, A.~Skarmeta,
  U.~Simsek, A.~Fensel, {Data Aggregation, Fusion and Recommendations for
  Strengthening Citizens Energy-aware Behavioural Profiles}, in: 2017 Glob.
  Internet Things Summit, 2017, pp. 1--6.

\bibitem{Zorrilla2019}
M.~Zorrilla, {\'{A}}.~Ibrain, {Bernard, an energy intelligent system for
  raising residential users awareness}, Comput. Ind. Eng. 135~(June) (2019)
  492--499.

\bibitem{Machorro-Cano2020}
I.~Machorro-Cano, G.~Alor-Hern{\'{a}}ndez, M.~A. Paredes-Valverde,
  L.~Rodr{\'{i}}guez-Mazahua, J.~L. S{\'{a}}nchez-Cervantes, J.~O.
  Olmedo-Aguirre, {HEMS-IoT: A Big Data and Machine Learning-Based Smart Home
  System for Energy Saving}, Energies 13~(5).

\bibitem{Sitoula2019}
P.~Sitoula, D.~Rahayu, P.~D. Haghighi, S.~Goodwin, C.~Ling, {Context-Aware
  Smart Energy Recommender (CASER)}, in: Proc. 17th Int. Conf. Adv. Mob.
  Comput. Multimed., 2019, pp. 13--19.

\bibitem{Alsalemi2020}
A.~Alsalemi, F.~Bensaali, A.~Amira, N.~Fetais, C.~Sardianos, I.~Varlamis,
  {Smart Energy Usage and Visualization based on Micro-Moments}, in: Adv.
  Intell. Syst. Comput., Vol. 1038, 2020, pp. 557--566.

\bibitem{Al-Kababji2020}
A.~Al-Kababji, A.~Alsalemi, Y.~Himeur, F.~Bensaali, A.~Amira, R.~Fernandez,
  N.~Fetais, Energy data visualizations on smartphones for triggering
  behavioral change: Novel vs. conventional, in: 2020 2nd Global Power, Energy
  and Communication Conference (GPECOM), IEEE, 2020, pp. 312--317.

\bibitem{al2020interactive}
A.~Al-Kababji, A.~Alsalemi, Y.~Himeur, F.~Bensaali, A.~Amira, R.~Fernandez,
  N.~Fetais, Interactive visual analytics for residential energy big data,
  Information Visualization (2021) 1--20.

\bibitem{BBC2020}
BBC, \href{https://www.bbc.com/news/technology-52331534}{{Coronavirus: Domestic
  electricity use up during day as nation works from home - BBC News}} (2020).
\url{https://www.bbc.com/news/technology-52331534}

\bibitem{Himeur2020IJIS-NILM}
Y.~{Himeur}, A.~{Elsalemi}, F.~{Bensaali}, A.~Amira, {An intelligent
  non-intrusive load monitoring scheme based on 2D phase encoding of power
  signals}, International Journal of Intelligent Systems 36~(1) (2021) 72--93.

\bibitem{siddiqui:hal-02954362}
A.~S. Siddiqui, A.~M. Sibal, {Energy Disaggregation in Smart Home Appliances: A
  Deep Learning Approach}, {Energy} (2020) 1--16.

\bibitem{himeur2020detection}
Y.~{Himeur}, A.~{Elsalemi}, F.~{Bensaali}, A.~Amira, Detection of
  appliance-level abnormal energy consumption in buildings using autoencoders
  and micro-moments, in: The Fifth International Conference on Big Data and
  Internet of Things (BDIoT), 2021, pp. 1--13.

\bibitem{himeur2020anomaly}
Y.~Himeur, K.~Ghanem, A.~Alsalemi, F.~Bensaali, A.~Amira, Artificial
  intelligence based anomaly detection of energy consumption in buildings: A
  review, current trends and new perspectives, Applied Energy 287 (2021)
  116601.

\bibitem{Himeur2020IJIS-AD}
Y.~{Himeur}, A.~{Elsalemi}, F.~{Bensaali}, A.~Amira, Smart power consumption
  abnormality detection in buildings using micro-moments and improved k-nearest
  neighbors, International Journal of Intelligent Systems (2021) 1--25.

\bibitem{gong2020edgerec}
Y.~Gong, Z.~Jiang, Y.~Feng, B.~Hu, K.~Zhao, Q.~Liu, W.~Ou, Edgerec: Recommender
  system on edge in mobile taobao, in: Proceedings of the 29th ACM
  International Conference on Information \& Knowledge Management, 2020, pp.
  2477--2484.

\bibitem{sun2020convergence}
C.~Sun, H.~Li, X.~Li, J.~Wen, Q.~Xiong, W.~Zhou, Convergence of recommender
  systems and edge computing: A comprehensive survey, IEEE Access 8 (2020)
  47118--47132.

\bibitem{su2019edge}
X.~Su, G.~Sperl{\`\i}, V.~Moscato, A.~Picariello, C.~Esposito, C.~Choi, An edge
  intelligence empowered recommender system enabling cultural heritage
  applications, IEEE Transactions on Industrial Informatics 15~(7) (2019)
  4266--4275.

\bibitem{himeur2020emergence}
Y.~{Himeur}, A.~{Elsalemi}, F.~{Bensaali}, A.~Amira, The emergence of hybrid
  edge-cloud computing for energy efficiency in buildings, in: Proceedings of
  SAI Intelligent Systems Conference, 2021, pp. 1--12.

\bibitem{hidasi2018cutting}
B.~Hidasi, Cutting-edge collaborative recommendation algorithms: Deep learning.
  (2018).

\bibitem{felfernig2017recommendation}
A.~Felfernig, S.~P. Erdeniz, M.~Jeran, A.~Akcay, P.~Azzoni, M.~Maiero,
  C.~Doukas, Recommendation technologies for iot edge devices., in:
  FNC/MobiSPC, 2017, pp. 504--509.

\bibitem{sayed2021endorsing}
A.~{Sayed}, A.~{Elsalemi}, Y.~{Himeur}, F.~{Bensaali}, A.~Amira, Endorsing
  energy efficiency through accurate appliance-level power monitoring,
  automation and data visualization, in: The 4th International Conference on
  Networking, Information Systems \& Security (NISS 2021), 2021, pp. 1--13.

\bibitem{wang2020privacy}
X.~Wang, B.~Gu, Y.~Qu, Y.~Ren, Y.~Xiang, L.~Gao, A privacy preserving
  aggregation scheme for fog-based recommender system, in: International
  Conference on Network and System Security, Springer, 2020, pp. 408--418.

\bibitem{zhou2020cost}
X.~Zhou, R.~Canady, S.~Bao, A.~Gokhale, Cost-effective hardware accelerator
  recommendation for edge computing, in: 3rd $\{$USENIX$\}$ Workshop on Hot
  Topics in Edge Computing (HotEdge 20), 2020.

\bibitem{ibrahim2020fog}
T.~S. Ibrahim, A.~I. Saleh, N.~Elgaml, M.~M. Abdelsalam, A fog based
  recommendation system for promoting the performance of e-learning
  environments, Computers \& Electrical Engineering 87 (2020) 106791.

\bibitem{jabeen2019iot}
F.~Jabeen, M.~Maqsood, M.~A. Ghazanfar, F.~Aadil, S.~Khan, M.~F. Khan,
  I.~Mehmood, An iot based efficient hybrid recommender system for
  cardiovascular disease, Peer-to-Peer Networking and Applications 12~(5)
  (2019) 1263--1276.

\end{thebibliography}
\end{document}